\documentclass[12pt,draftcls,journal,onecolumn]{IEEEtran}

\usepackage{graphicx}
\usepackage[cmex10]{amsmath}
\usepackage{amsfonts}
\usepackage{amssymb}
\usepackage{mathrsfs}
\usepackage{algorithmic}
\usepackage{algorithm}
\usepackage{array}
\usepackage{mdwmath}
\usepackage{mdwtab}
\usepackage[caption=false,font=footnotesize]{subfig}
\usepackage{fixltx2e}
\usepackage{stfloats}
\usepackage{footnote}
\usepackage{tabularx}
\usepackage{multirow}
\usepackage{color}
\usepackage{textcomp}

\newtheorem{definition}{Definition}
\newtheorem{proposition}{Proposition}
\newtheorem{theorem}{Theorem}
\newtheorem{lemma}{Lemma}

\newtheorem{remark}{Remark}
\newtheorem{assumption}{Assumption}
\newtheorem{example}{Example}

\begin{document}

\title{Repeated Games With Intervention: \\Theory and Applications in Communications}

\author{Yuanzhang~Xiao, Jaeok~Park, and~Mihaela~van~der~Schaar%
\thanks{Y.~Xiao and M.~van der Schaar are with Department of Electrical Engineering, UCLA. Email: \{yxiao,mihaela\}@ee.ucla.edu.}
\thanks{J.~Park was with Department of Economics and Department of Electrical Engineering, UCLA, and is now with School of Economics, Yonsei
University, Seoul, ROK. Email: jaeok.park@yonsei.ac.kr.}}

\maketitle

\begin{abstract}
In communication systems where users share common resources, users' selfish behavior usually results in suboptimal resource utilization. There have
been extensive works that model communication systems with selfish users as one-shot games and propose incentive schemes to achieve Pareto optimal
action profiles as non-cooperative equilibria. However, in many communication systems, due to strong negative externalities among users, the sets of
feasible payoffs in one-shot games are nonconvex. Thus, it is possible to expand the set of feasible payoffs by having users choose convex
combinations of different payoffs. In this paper, we propose a repeated game model generalized by intervention. First, we use repeated games to
convexify the set of feasible payoffs in one-shot games. Second, we combine conventional repeated games with intervention, originally proposed for
one-shot games, to achieve a larger set of equilibrium payoffs and loosen requirements for users' patience to achieve it. We study the problem of
maximizing a welfare function defined on users' equilibrium payoffs, subject to minimum payoff guarantees. Given the optimal equilibrium payoff, we
derive the minimum intervention capability required and design corresponding equilibrium strategies. The proposed generalized repeated game model
applies to various communication systems, such as power control and flow control.
\end{abstract}


\section{Introduction}
Game theory is a formal framework to model and analyze the interactions of selfish agents. It has been used in the literature to study communication
networks with selfish agents \cite{MacKenzieWicker01}\cite{Altman_Survey06}. Most works modeled communication systems as one-shot games, studied the
inefficiency of noncooperative outcomes, and proposed incentive schemes, such as pricing and auctions \cite{HuangBerry_JSAC06}--\cite{ShenBasar07},
to improve the inefficient outcomes towards the Pareto boundary.

Recently, a new incentive scheme, called ``intervention'', has been proposed \cite{ParkMihaela_JSAC}, with applications to medium access control
(MAC) games \cite{ParkMihaela_EURASIP}\cite{ParkMihaela_Gamenets} and power control games \cite{XiaoMihaela_JSTSP}.\footnote{With the same philosophy
as intervention, a packet-dropping incentive scheme was proposed for flow control games in \cite{GaiKrishnamachari_Infocom}.} In an intervention
scheme, the designer places an intervention device, which has a monitoring technology to monitor the user behavior and an intervention capability to
intervene in their interaction, in the system. The intervention device observes a signal about the actions of agents, and chooses an intervention
action depending on the observed signal. In this way, it can punish misbehavior of an agent by exerting intervention following a signal that suggests
a deviation. One of the advantages of intervention is that the intervention device directly interacts with the users in the system, instead of using
outside instruments such as monetary payments as pricing and auctions do. As a result, intervention can provide more robust incentives in the sense
that agents cannot avoid intervention. Moreover, in contrast to pricing and auctions, intervention requires no knowledge on the users' valuation of
the resource usage in some scenarios \cite{XiaoMihaela_JSTSP}.

In some communication systems where users create severe congestion or interference, increasing a user's payoff requires a significant sacrifice of
others' payoffs. This feature is reflected by a nonconvex set of feasible payoffs in some systems studied in the aforementioned works
\cite{HuangBerry_JSAC06}--\cite{GaiKrishnamachari_Infocom} using one-shot game models. For example, in one-shot power control games, the set of
feasible payoffs is nonconvex when the cross channel gains are large \cite{StanczakBoche_IT07}\cite{LarssonJorswieck09}. In one-shot MAC games based
on the collision model, the set of feasible payoffs is also nonconvex, because transmissions from multiple users cause packet loss
\cite{YangKimZhangChiangTan_INFOCOM11}\cite{ParkMihaela_ToN}. Moreover, we will see in this paper that the sets of feasible payoffs of some one-shot
flow control games are also nonconvex. To sum up, the sets of feasible payoffs are nonconvex in many communication scenarios, and when the set of
feasible payoffs is nonconvex, its Pareto boundary can be dominated by a convex combination of different payoffs. In one-shot games, such convex
combinations cannot be achieved unless a public correlation device is used.\footnote{Public correlation devices are used in game theory literature to
simplify the construction of the payoffs that are convex combinations of pure-action payoffs. Such devices may not be available in communication
networks. Even if there exist such devices, there are additional costs on broadcasting the random signals generated by public correlation devices.}

Although some works in power control \cite{StanczakBoche_IT07}\cite{LarssonJorswieck09} and medium access control \cite{ParkMihaela_ToN} proposed
time-sharing solutions that achieve payoffs beyond the set of feasible payoffs in one-shot games, they did not consider users' incentives.
Specifically, in a time-sharing protocol, a user may transmit at the time slots that are assigned to other users, in order to obtain a higher payoff.
Hence, it is important to study deviation-proof protocols, which make it in the self-interest of users to comply with the protocols. To this end, we
use repeated games to model the communication scenarios. In a repeated game, a stage game is played repeatedly, and a user's payoff in the repeated
game is the discounted average or the limit of the mean of the stage-game payoffs. Users can choose different actions in the stage games in different
periods, resulting in a convex combination of different stage-game payoffs as the repeated game payoff. A repeated game strategy prescribes what
action to take given past observations and thus can be interpreted as a protocol. If a repeated game strategy constitutes a subgame perfect
equilibrium (SPE), then no user can gain from deviation at any occasion. Hence, a SPE strategy can be considered as a deviation-proof protocol.

In this paper, we consider the protocol design problem of maximizing some welfare function defined on users' (subgame perfect) equilibrium payoffs,
subject to minimum payoff guarantees for all users. When we design a protocol in the repeated game framework, there are three important
considerations, which also motivates us to introduce intervention in a repeated game. The first one is the set of equilibrium payoffs, which
characterizes payoffs that can be achieved at an SPE. Since the designer is optimizing some welfare function, for example, the sum or the minimum of
all users' payoffs, this set, along with the minimum payoff guarantees, determines the feasible set of the optimization problem. Consequently, a
larger set of equilibrium payoffs can result in higher social welfare. In this paper, we will characterize the set of equilibrium payoffs in repeated
games with intervention, and show that using intervention can yield a larger set of equilibrium payoffs than the corresponding set without
intervention.

\begin{figure}
\centering
\includegraphics[width =4.0in]{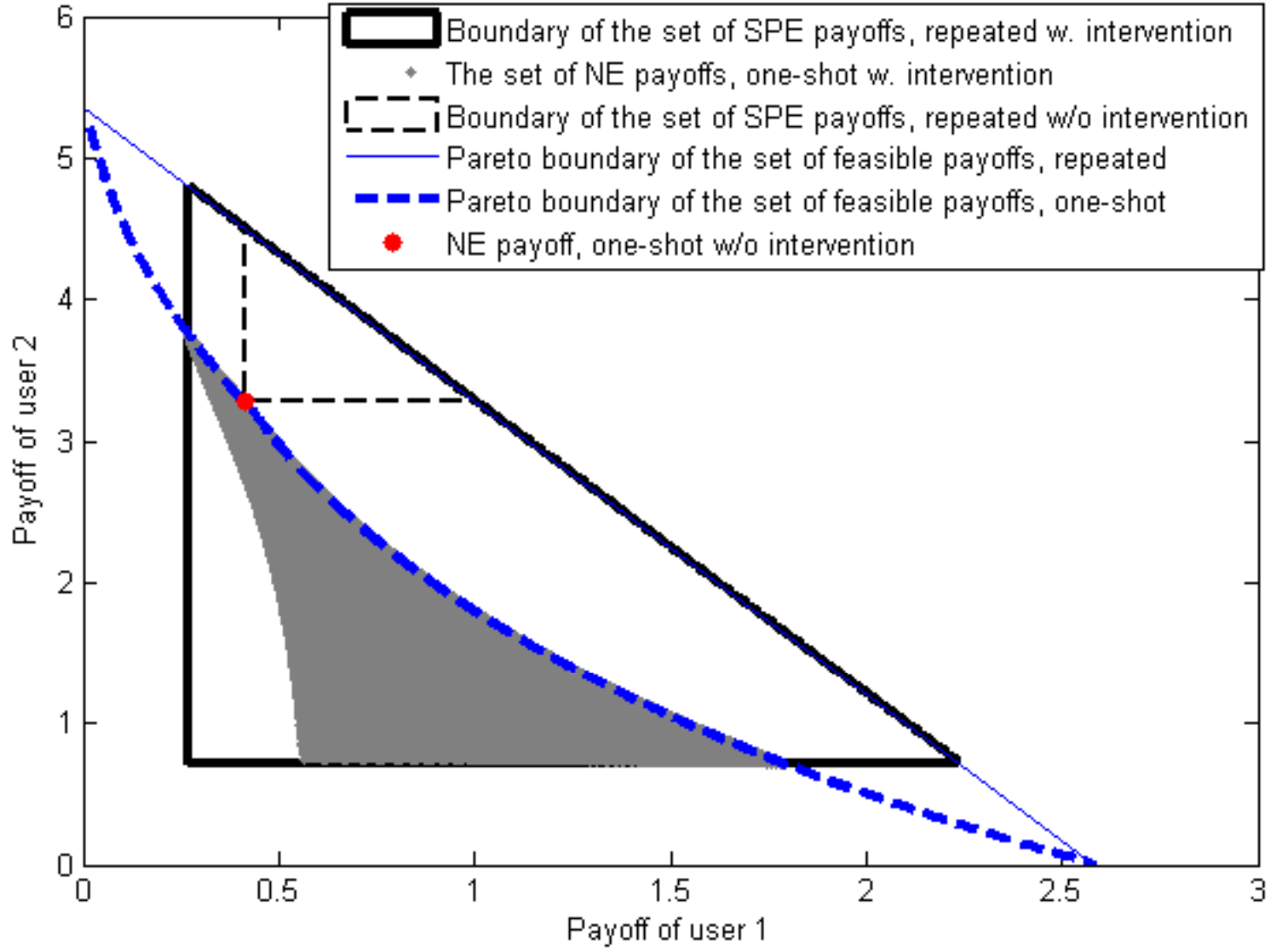}
\caption{Payoffs in the one-shot game model and the repeated game model with and without intervention for an example two-user system. The dashed
curve is the Pareto boundary of the set of feasible payoffs in one-shot games. The solid line on the upper right is the Pareto boundary of the set of
feasible payoffs in repeated games. The gray area is the set of Nash equilibrium payoffs in one-shot games with intervention. The sets of subgame
perfect equilibrium payoffs in repeated games with and without intervention are within the boundaries shown in the figure(the discount factor
approaches $1$ in the cases of repeated games). The dot is the Nash equilibrium payoff of the one-shot game or of the stage game of the repeated
game.} \label{fig:UtilityRegion_PowerControl_2user_OneshotRepeatedIntervention_AllInOne}
\end{figure}

An illustration of the promising gain by using repeated games with intervention is shown in
Fig.~\ref{fig:UtilityRegion_PowerControl_2user_OneshotRepeatedIntervention_AllInOne}. We plot the set of equilibrium payoffs in a two-user flow
control game\footnote{Although we use flow control as an example, the qualitative result is true for other scenarios, such as power control and
medium access control.} under different incentive schemes. We use the same intervention device (thus the same intervention capability) in the games
with intervention. We can see that the set of equilibrium payoffs of the repeated game with intervention includes the set of equilibrium payoffs in
the one-shot game with intervention and the set of equilibrium payoffs in the repeated game without intervention.

The second consideration is the discount factor, which is affected by the user patience and the network dynamics. The discount factor represents the
rate at which users discount future payoffs; a more patient user has a larger discount factor. The discount factor can also model the probability of
users remaining in the network in each period; a more dynamic network results in a smaller discount factor. As we have mentioned above, the designer
aims to maximize some welfare function on the feasible set determined by the set of equilibrium payoffs and minimum payoff guarantees. Given the
target payoff that is in the feasible set and maximizes the welfare function, there is a minimum requirement on the discount factors to achieve it as
an SPE payoff. A lower discount factor is desirable in the sense that with a lower requirement, a protocol is effective in a wider variety of users
and more dynamic networks. In this paper, we will determine the minimum requirement on the discount factor to support the target payoff, and show
that using intervention can lower the minimum requirement compared to the case without intervention. Moreover, we obtain a trade-off between the
discount factor, the minimum payoff guarantees, and the intervention capability. Hence, given a discount factor, we can calculate the minimum
intervention capability required to support the target payoff. Conversely, given the intervention capability available, we can calculate the minimum
requirement on the discount factor, and thus determine the types of users and networks that can be supported.

The last consideration is the equilibrium strategy. Given a target payoff and the discount factor, we show how to construct a candidate equilibrium
strategy, namely the deviation-proof protocol. We will also see that intervention can simplify the users' equilibrium strategies, which reduces the
complexity of users' devices to execute a protocol.

To the best of our knowledge, our paper is the first one to study repeated games with intervention systematically, addressing the above three
considerations. The rest of this paper is organized as follows. Section~\ref{sec:SystemModel} describes a repeated game model generalized by
intervention and formulates a protocol design problem. In Section~\ref{sec:EquilibriumPayoffSet}, we characterize the set of equilibrium payoffs when
the discount factor is close to one, and specify the structure of equilibrium strategies. Then we analyze the design problem in details in
Section~\ref{sec:DesignProblem}. Simulation results are presented in Section~\ref{sec:Simulation}. Finally, Section~\ref{sec:Conclusion} concludes
the paper.

\section{Model of Repeated Games With Intervention}\label{sec:SystemModel}

\subsection{The Stage Game With Intervention}
We consider a system with $N$ users. The set of the users is denoted by $\mathcal{N} \triangleq \{1,2,\ldots,N\}$. Each user $i$ chooses its
action\footnote{We consider only pure actions in this paper. Hence, we use ``action'' to mean ``pure action'', and use ``pure action'' only when we
want to emphasize it.} $a_i$ from the set $A_i\subset\mathbb{R}^{k_i}$ for some integer $k_i>0$. The set of action profiles is denoted by
$\mathcal{A}=\prod_{i=1}^N A_i$, and the action profile of all the users is denoted by ${\bf a}=(a_1,\ldots,a_N) \in \mathcal{A}$. Let ${\bf a}_{-i}$
be the action profile of all the users other than user $i$. In addition to the $N$ users, there exists an intervention device in the network, indexed
by $0$. The intervention device chooses its action $a_0$ from the set $A_0\subset\mathbb{R}^{k_0}$ for some integer $k_0>0$. We call the set $A_0$
the intervention capability (of the intervention device), because it determines the actions that the intervention device can take when it intervenes
in the interaction among users. The payoffs of the users are determined by the actions of the users and the intervention device, and user $i$'s
payoff function is denoted by $u_i:A_0\times\mathcal{A}\rightarrow\mathbb{R}$.

We assume that there exists a null intervention action, denoted by $\underline{a}_0 \in A_0$, which corresponds to the case where there is no
intervention device. We further assume that an intervention action can only decrease the payoffs of the users, i.e., $u_i(a_0, \mathbf{a}) \leq
u_i(\underline{a}_0, \mathbf{a})$ for all $a_0 \in A_0$, all $\mathbf{a} \in \mathcal{A}$, and all $i$. In other words, intervention can provide only
punishment to users, not rewards.

An important feature of the intervention device is that it does not have its own objective and can be programmed in the way that the protocol
designer desires. Hence, the Nash equilibrium $(a_0^*,\mathbf{a}^*)$ of the stage game with intervention is defined by
\begin{eqnarray}\label{eqn:DefinitionNashEquilibrium}
u_i(a_0^*,\mathbf{a}^*)\geq u_i(a_0^*,a_i,\mathbf{a}_{-i}^*),~\forall i\in\mathcal{N},~\forall a_i\in A_i.
\end{eqnarray}
The Nash equilibrium $(\underline{a}_0,\mathbf{a}^*)$ without intervention is defined similarly by fixing $a_0^*=\underline{a}_0$ in
\eqref{eqn:DefinitionNashEquilibrium}.

%\vspace{-.5cm}
\subsection{The Repeated Game With Intervention}
In the repeated game, the stage game is played in every period $t=0,1,2,\ldots$. At the end of period $t$, all the users and the intervention device
observe the action profile at period $t$, denoted by $(a_0^t,\mathbf{a}^t)$. That is, we assume perfect monitoring. As a result, the users and the
intervention device share the same history at each period, and the history at period $t$ is the collection of all the actions taken before period
$t$. We denote the history at period $t\geq1$ by $h^t=(a_0^0,\mathbf{a}^0;a_0^1,\mathbf{a}^1;\ldots;a_0^{t-1},\mathbf{a}^{t-1})$. The history at
period 0 is set as $h^0=\emptyset$. The set of possible histories at period $t$ is denoted by $\mathscr{H}^t$, and the set of all possible histories
by $\mathscr{H}=\bigcup_{t=0}^{\infty} \mathscr{H}^t$.

The (pure) strategy of user $i$ is a mapping from the set of all possible histories to its action set, written as $\sigma_i:\mathscr{H}\rightarrow
A_i,~i\in\mathcal{N}$. User $i$'s action at history $h^t$ is determined by $a_i^t=\sigma_i(h^t)$. The joint strategy of all the users is
$\mathbf{\sigma}=(\sigma_1,\ldots,\sigma_N)$, and the joint strategy of all the users other than user $i$ is $\mathbf{\sigma}_{-i}$. The joint action
at history $h^t$ is $\mathbf{a}^t=\mathbf{\sigma}(h^t)$. The action of the intervention device at history $h^t$ is determined by
$a_0^t=\sigma_0(h^t)$, where $\sigma_0:\mathscr{H}\rightarrow A_0$ is the intervention rule.\footnote{Note that we do not use ``intervention
strategy'' here, because the intervention device is not a strategic player and just follows a rule prescribed by the designer.} When the intervention
rule is constant at $\underline{a}_0$, namely $\sigma_0(h)=\underline{a}_0$ for all $h\in\mathscr{H}$, the repeated game with intervention reduces to
the conventional repeated game without intervention.

The overall payoff is the normalized sum of discounted payoffs at each period. We assume all the users have the same discount factor
$\delta\in[0,1)$. Then the payoff function of user $i$ in the repeated game is
\begin{eqnarray}\label{DiscountedPayoff_Simul}
\begin{array}{c}U_i(\sigma_0,\mathbf{\sigma}) = (1-\delta) \sum_{t=0}^\infty \delta^t
u_i(a_0^t(\sigma_0,\sigma),\mathbf{a}^t(\sigma_0,\sigma)),\end{array}
\end{eqnarray}
where $(a_0^t(\sigma_0,\sigma),\mathbf{a}^t(\sigma_0,\sigma))$ is the $t$th-period actions of the intervention device and the users induced by the
intervention rule $\sigma_0$ and the joint strategy $\sigma$. $(a_0^t(\sigma_0,\sigma),\mathbf{a}^t(\sigma_0,\sigma))$ can be calculated recursively
as
\begin{eqnarray}
(a_0^t(\sigma_0,\sigma),\mathbf{a}^t(\sigma_0,\sigma))&=&\left(\sigma_0(a_0^0(\sigma_0,\sigma),\mathbf{a}^0(\sigma_0,\sigma);\ldots;a_0^{t-1}(\sigma_0,\sigma),\mathbf{a}^{t-1}(\sigma_0,\sigma)),\right. \nonumber \\
& &~\left.\sigma(a_0^0(\sigma_0,\sigma),\mathbf{a}^0(\sigma_0,\sigma);\ldots;a_0^{t-1}(\sigma_0,\sigma),\mathbf{a}^{t-1}(\sigma_0,\sigma))\right)
\end{eqnarray}

User $i$'s continuation strategy induced by any history $h^t\in\mathscr{H}$, denoted $\sigma_i|_{h^t}$, is defined by
$\sigma_i|_{h^t}(h^\tau)=\sigma_i(h^t h^\tau), \forall h^\tau \in \mathscr{H}$, where $h^t h^\tau$ is the concatenation of the history $h^t$ followed
by the history $h^\tau$. Similarly, we can define $\sigma_0|_{h^t}$ for the intervention device. By convention, we denote $\sigma|_{h^t}$ and
$\sigma_{-i}|_{h^t}$ the strategy profile of all the users and the strategy profile of all the users other than user $i$, induced by $h^t$,
respectively. Then the subgame perfect equilibrium of the repeated game is the intervention rule and the strategy profile $(\sigma_0,\sigma)$ that
satisfies
\begin{eqnarray}
U_i(\sigma_0|_{h^t},\sigma|_{h^t}) \geq
U_i(\sigma_0|_{h^t},\sigma_i'|_{h^t},\sigma_{-i}|_{h^t}),~\mathrm{for~all}~\sigma_i',~\mathrm{~for~all}~i\in\mathcal{N},~\mathrm{and~for~all}~h^t\in\mathscr{H}.
\end{eqnarray}
The subgame perfect equilibrium prescribes a strategy profile from which no user has incentive to deviate at any period and at any history. Hence,
the equilibrium strategy can be considered as a deviation-proof protocol.

%\vspace{-.5cm}
\subsection{Problem Formulation}
There is a protocol designer who chooses an intervention rule $\sigma_0$ and recommends the joint strategy $\sigma$ to the users. We assume that the
designer knows the structure of the game including the number of users, action spaces, and payoff functions. The designer maximizes a welfare
function defined over the repeated-game payoffs of the users, $W(U_1(\sigma_0,\sigma), \ldots, U_N(\sigma_0,\sigma))$. At the maximum of the welfare
function, some users may have low payoffs. To avoid this, the designer provides a minimum payoff guarantee $\gamma_i$ for each user $i$. Hence, we
can formally define the protocol design problem as follows
\begin{eqnarray}\label{eqn:ProtocolDesignProblem}
&\displaystyle\max_{\sigma_0,\sigma}& W(U_1(\sigma_0,\sigma), \ldots, U_N(\sigma_0,\sigma)) \\
&s.t.& (\sigma_0, \sigma)~\mathrm{is~subgame~perfect~equilibrium}, \nonumber\\
&    & U_i(\sigma_0,\sigma)\geq \gamma_i,~\forall i\in\mathcal{N}. \nonumber
\end{eqnarray}

Examples of welfare functions are the sum payoff $\sum_{i=1}^N U_i$, and the absolute fairness $\min_{i\in\mathcal{N}} U_i$. Note that the first step
towards solving the design problem is to characterize the set of equilibrium payoffs, which will be the focus of the next section. Once the designer
identifies what payoffs are achievable as SPE, it can maximize $W$ over the obtained set of payoffs satisfying the constraints.

\section{The Set of Equilibrium Payoffs And Structures of Equilibrium Strategies}\label{sec:EquilibriumPayoffSet}
In this section, we determine the set of equilibrium payoffs for repeated games with intervention, when the discount factor is sufficiently close to
$1$. For the protocol designer, it is important to recognize which payoffs are achievable at SPE. The set of equilibrium payoffs in conventional
repeated games, when the discount factor is sufficiently close to $1$, is characterized by folk theorems. In this section, we adapt conventional folk
theorems for repeated games with intervention. Our proofs are constructive and thus yield structures of the equilibrium strategies. Our results show
that intervention can enlarge the set of equilibrium payoffs and enable users to use simpler strategies while sufficient conditions for folk theorems
are still satisfied.

To state folk theorems, we need to define pure-action minmax payoffs with intervention.
\begin{definition}[Pure-action Minmax Payoff With Intervention]
User $i$'s pure-action minmax payoff with intervention is defined as
\begin{eqnarray}\label{eqn:MinmaxPayoff_WithIntervention}
\underline{v}_i^w \triangleq \min_{(a_0,\mathbf{a}_{-i}) \in A_0 \times \mathcal{A}_{-i}} \max_{a_i \in A_i} u_i(a_0, a_i, \mathbf{a}_{-i}).
\end{eqnarray}
\end{definition}

We say a payoff $\mathbf{v}$ is strictly individually rational, if $v_i>\underline{v}_i^w$ for all $i\in\mathcal{N}$. Then the set of feasible and
strictly individually rational payoffs can be written as
\begin{eqnarray}\label{eqn:IndividuallyRationalPayoffSet_WithIntervention}
\mathscr{V}_w^{\dag p} = \{ \mathbf{v} \in \mathscr{V}_w^{\dag} : v_i > \underline{v}_i^w,~\forall i\in\mathcal{N} \},
\end{eqnarray}
where $\mathscr{V}_w^{\dag}$ is the set of feasible payoffs, defined by
\begin{eqnarray}\label{eqn:FeasiblePayoffSet_WithIntervention}
\mathscr{V}_w^{\dag} = \mathrm{co} \left\{ \mathbf{v} \in \mathbb{R}^N: \exists (a_0,\mathbf{a}) \in A_0 \times \mathcal{A},~\mathrm{s.t.}~\mathbf{v}
= u(a_0, \mathbf{a}) \right\},
\end{eqnarray}
where $\mathrm{co} X$ denotes the convex hull of a set $X$.

In the rest of this section, we will prove two folk theorems, depending on if there exists a mutual minmax profile.

%\vspace{-.5cm}
\subsection{Games With Mutual Minmax Profiles}
First, we prove the folk theorem for the games that have mutual minmax profiles. Before we state the folk theorem, we define the mutual minmax
profile for repeated games with intervention.
\begin{definition}[Mutual Minmax Profile]\label{definition:MutualMinmaxProfile}
An action profile $(\hat{a}_0,\mathbf{\hat{a}})$ is a mutual minmax profile if it satisfies $\underline{v}_i^w = \min_{a_0,\mathbf{a}_{-i}}
\max_{a_i} u_i(a_0,a_i,\mathbf{a}_{-i}) = \max_{a_i} u_i(\hat{a}_0,a_i,\mathbf{\hat{a}}_{-i}),~\forall i\in\mathcal{N}$.
\end{definition}

Now we can state the minmax folk theorem for repeated games with intervention as follows.
\begin{proposition}[Minmax Folk Theorem]\label{proposition:FolkTheorem_Minmax_PureAction}
Suppose that there exists a mutual minmax profile $(\hat{a}_0,\mathbf{\hat{a}})$. Then for every feasible and strictly individually rational payoff
$\mathbf{v}\in{\mathscr{V}}_w^{\dag p}$, there exists $\underline{\delta}<1$ such that for all $\delta\in(\underline{\delta},1)$, there exists a
subgame perfect equilibrium with payoffs $\mathbf{v}$, of the repeated game with intervention.
\end{proposition}
\begin{IEEEproof}
See \cite[Appendix~A]{XiaoParkMihaela_Appendix}.
\end{IEEEproof}

\subsubsection{Structure of the equilibrium strategy}
We first briefly describe the structure of the equilibrium strategy. Then we formally present the equilibrium strategy as an automaton
\cite{HopcroftMotwaniUllman}.

Suppose we want to achieve a SPE payoff $\mathbf{v}$. When the discount factor is sufficiently close to $1$, there exists a sequence of action
profiles $\{\tilde{a}_0^\tau,\mathbf{\tilde{a}}^\tau\}_{\tau=0}^{T-1}$ for some integer $T>0$, which satisfies
\begin{eqnarray}
\frac{1-\delta}{1-\delta^T}\cdot\sum_{\tau=0}^{T-1} \delta^\tau \mathbf{u}(\tilde{a}_0^\tau,\mathbf{\tilde{a}}^\tau) = \mathbf{v}.
\end{eqnarray}
Note that $T$ is infinite in general. When $T=\infty$, the sequence $\{\tilde{a}_0^\tau,\mathbf{\tilde{a}}^\tau\}_{\tau=0}^\infty$ yields the desired
payoff $(1-\delta)\cdot\sum_{\tau=0}^\infty \delta^\tau \mathbf{u}(\tilde{a}_0^\tau,\mathbf{\tilde{a}}^\tau)=\mathbf{v}$. When $T$ is finite,
repeating the sequence $\{\tilde{a}_0^\tau,\mathbf{\tilde{a}}^\tau\}_{\tau=0}^{T-1}$ forever yields the desired payoff
\begin{eqnarray}
(1-\delta)\cdot\left(\sum_{\tau=0}^{T-1} \delta^\tau \mathbf{u}(\tilde{a}_0^\tau,\mathbf{\tilde{a}}^\tau)+\delta^T\cdot\sum_{\tau=0}^{T-1}
\delta^\tau \mathbf{u}(\tilde{a}_0^\tau,\mathbf{\tilde{a}}^\tau)+\cdots\right)=\frac{1-\delta}{1-\delta^T}\cdot\sum_{\tau=0}^{T-1} \delta^\tau
\mathbf{u}(\tilde{a}_0^\tau,\mathbf{\tilde{a}}^\tau)=\mathbf{v}.
\end{eqnarray}

In the equilibrium strategy, the intervention device and the users start from $(\tilde{a}_0^0,\mathbf{\tilde{a}}^0)$ at $t=0$, and follow the
sequence $\{\tilde{a}_0^\tau,\mathbf{\tilde{a}}^\tau\}_{\tau=1}^{T-1}$ afterwards. If $T$ is finite, they repeat this sequence forever. Since the
sequence $\{\tilde{a}_0^\tau,\mathbf{\tilde{a}}^\tau\}_{\tau=0}^{T-1}$ is played at the equilibrium, it is also called the equilibrium outcome path.
When a deviation from user $i$ happens, the intervention device plays $\hat{a}_0$, and the other users play $\mathbf{\hat{a}}_{-i}$. The minmaxing
action $\hat{a}_0$ of the intervention device and the minmaxing action profile $\mathbf{\hat{a}}_{-i}$ of the users other than user $i$ last for $L$
periods as punishment. After the $L$ periods of punishment, the intervention device and the users return to the equilibrium outcome path. Any
deviation in the $L$ periods of punishment will trigger another new $L$ periods of punishment for the deviating user. The equilibrium strategy can be
described by the automaton in Fig.~\ref{fig:Automata_RepeatedGames_NotNE}.

\subsubsection{Discussions and examples}
From Proposition~\ref{proposition:FolkTheorem_Minmax_PureAction}, we can see that any payoff that strictly dominates the minmax payoff can be
achieved as a subgame perfect equilibrium payoff for some discount factors in repeated games with intervention. Here the impact of intervention is
two-fold. First, intervention can decrease the minmax payoff, which enlarges the set of equilibrium payoffs. Second, intervention may provide a
mutual minmax profile that is a Nash equilibrium for the stage game, while the mutual minmax profile in the original stage game without intervention
may not be a Nash equilibrium. If the mutual minmax profile is a Nash equilibrium, the punishment can be playing the mutual minmax profile forever
regardless of which user deviated, and the users cannot deviate from this severe punishment. On the other hand, for the mutual minmax profile that is
not a Nash equilibrium, we can use it as punishment for only a finite number of periods and use the promise of returning to the equilibrium outcome
path to deter users from deviating in the punishment phase. The latter punishment is not as strong as the previous one, and is more complicated in
terms of the associated automaton and the punishment length $L$ to be chosen properly.

\begin{example}[Power Control]
Now we consider a power control game to illustrate how intervention enlarges the set of equilibrium payoffs by decreasing the minmax payoff.

Consider a network with $N$ users transmitting power in a wireless channel. Use $i$'s action is its transmit power $a_i\in A_i=[0,\bar{a}_i]$. The
intervention device also transmits power $a_0\in A_0=[0,\bar{a}_0]$. User $i$'s signal-to-interference-and-noise ratio (SINR) is calculated by
$\frac{h_{ii} a_i}{h_{i0}a_0+\sum_{j\neq i} h_{ij}a_j + n_i}$, where $h_{ij}$ is the channel gain from user $j$'s transmitter to user $i$'s receiver,
and $n_i$ is the noise power at user $i$'s receiver. Each user $i$'s stage-game payoff is its throughput
\cite{AlpcanBasar_CDMA}\cite{LarssonJorswieck09}\cite{Chiang_FoundationTrend08}, namely
\begin{eqnarray}
u_i(a_0,a_i,\mathbf{a}_{-i}) = \log_2\left(1+\frac{h_{ii} a_i}{h_{i0}a_0+\sum_{j\neq i} h_{ij}a_j + n_i}\right).
\end{eqnarray}
Note that the payoff function can be an arbitrary increasing function of the SINR without changing the following analysis.

In this power control game, the null intervention action, which corresponds to the case with no intervention, is $\underline{a}_0=0$. Without
intervention (i.e., $a_0$ is fixed at $\underline{a}_0$), the only Nash equilibrium of the stage game is
$(\underline{a}_0,\mathbf{\bar{a}})=(0,\bar{a}_1,\ldots,\bar{a}_N)$, where every user transmits at its maximum power. Moreover, the Nash equilibrium
is the mutual minmax profile with payoff $\mathbf{\underline{v}}^o = (\underline{v}_1^o,\ldots,\underline{v}_N^o) = \mathbf{u}(0,\mathbf{\bar{a}})$.
With intervention, the mutual minmax profile $(\bar{a}_0,\mathbf{\bar{a}})$ is also a Nash equilibrium, with payoff $\mathbf{\underline{v}}^w =
(\underline{v}_1^w,\ldots,\underline{v}_N^w) = \mathbf{u}(\bar{a}_0,\mathbf{\bar{a}})<\mathbf{u}(0,\mathbf{\bar{a}}).$ Note that
$\underline{v}_i^w<\underline{v}_i^o~\forall i$, and that each $\underline{v}_i^w$ reduces as $\bar{a}_0$ increases. Hence, the set of equilibrium
payoffs when the discount factor is sufficiently close to $1$ expands as the maximum intervention power $\bar{a}_0$ increases.

%Fig.~\ref{fig:UtilityRegion_PowerControl_2user} confirms that the set of feasible and individually rational payoffs becomes larger as the maximum
%intervention power increases. Without intervention, there is no feasible and strictly individually rational payoff that achieves absolute fairness,
%at which both users have the same payoff. However, with intervention, the set of equilibrium payoffs is enlarged, and absolute fairness can be
%achieved. Also, as the maximum intervention power increases, we can achieve a larger total throughput at an equilibrium payoff.
\end{example}

\begin{figure}
\centering
\includegraphics[width =4.0in]{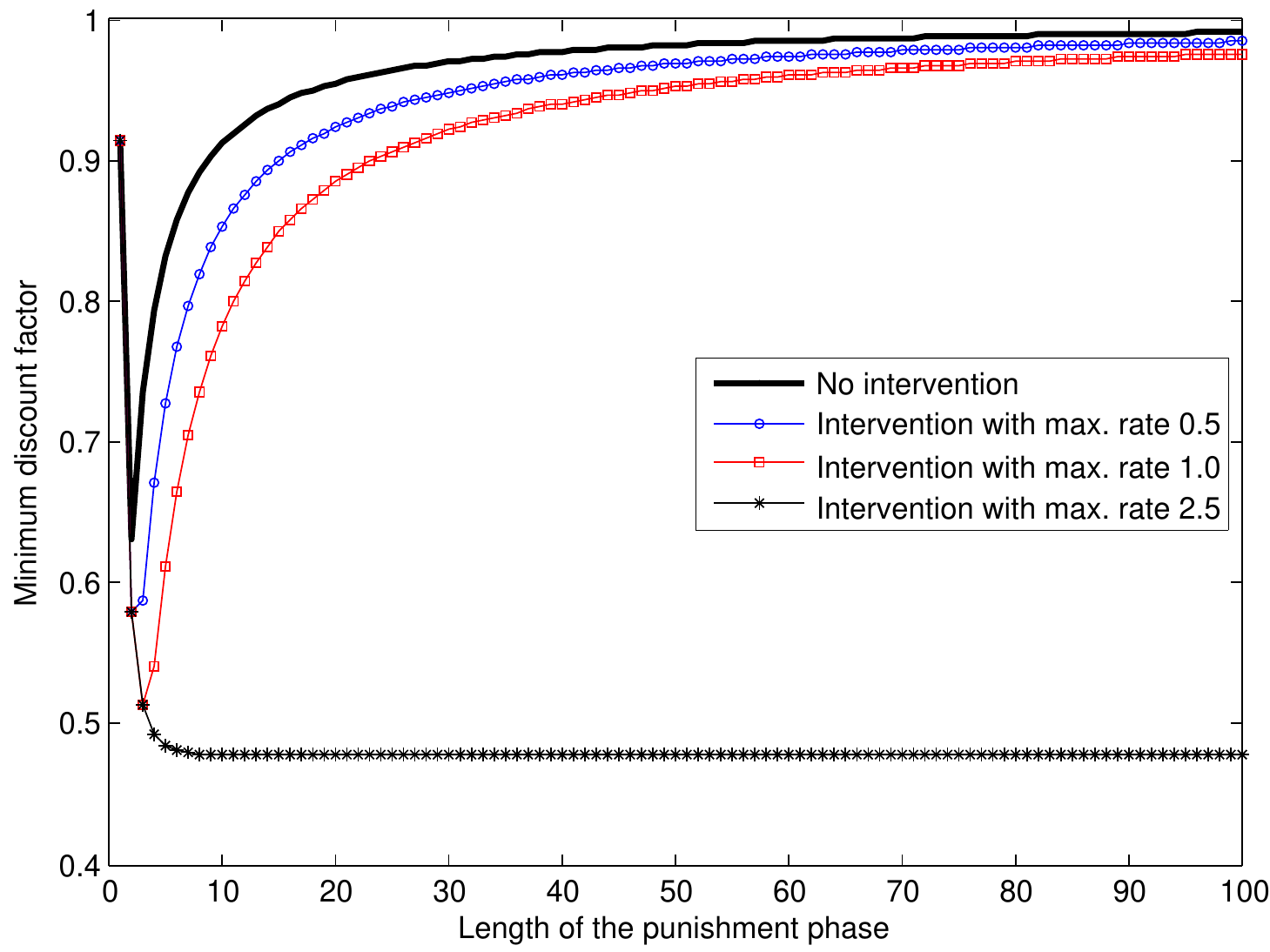}
\caption{Minimum discount factors to support the pure-action profile that achieves maximum sum payoff, under different punishment lengths and maximum
intervention flow rates. $N=4$. The service rate is $\mu=10$ bits/s. The maximum flow rates for all the users are $2.5$ bits/s. The trade-off factors
are $\beta_1=\beta_2=2$ and $\beta_3=\beta_4=3$.} \label{fig:MinDiscountFactor_PunishmentLength_FlowControl_color}
\end{figure}

\begin{example}
Now we consider a flow control game, whose mutual minmax profile without intervention may not be a Nash equilibrium. We show that intervention
induces a mutual minmax profile that is a Nash equilibrium, and thus, enables a more severe punishment and a simpler equilibrium strategy.

Consider a network with $N$ users transmitting packets through a single server, which can be modeled as an M/M/1 queue
\cite{BharathKumarJaffe}--\cite{ZhangDouligeris_TCOM}. User $i$'s action is its transmission rate $a_i\in A_i=[0,\bar{a}_i]$. The intervention device
also transmits packets at the rate of $a_0\in A_0=[0,\bar{a}_0]$. User $i$'s payoff is a function of its transmission rate $a_i$ and its delay
$1/(\mu-a_0-\sum_{j=1}^N a_j)$ \cite{YiMihaela_TCOM}--\cite{ZhangDouligeris_TCOM}, defined by
\begin{eqnarray}
\begin{array}{c}u_i(a_0,a_i,\mathbf{a}_{-i}) = a_i^{\beta_i}\max\left\{0,~\mu-a_0-\sum_{j=1}^N a_j\right\},\end{array}
\end{eqnarray}
where $\mu>0$ is the server's service rate, and $\beta_i>0$ is the parameter reflecting the trade-off between the transmission rate and the delay.
Here the ``max'' function indicates the fact that the payoff is zero when the total arrival rate is larger than the service rate. We assume that the
service rate is no smaller than the maximum total arrival rate without intervention, i.e., $\mu\geq\sum_{j=1}^N \bar{a}_j$.

Without intervention, the mutual minmax profile is every user transmitting at the maximum rate, i.e., $(0,\bar{a}_1,\ldots,\bar{a}_N)$. Since user
$i$'s best response to the action profile $\mathbf{a}_{-i}$ is given by $a_i^* = \min\left\{\frac{\beta_i}{1+\beta_i}\left(\mu-\sum_{j\neq i}
a_j\right),~\bar{a}_i\right\}$, the mutual minmax profile is a Nash equilibrium without intervention if and only if
\begin{eqnarray}
\begin{array}{c}\bar{a}_i \leq \frac{\beta_i}{1+\beta_i}\left(\mu-\sum_{j\neq i} \bar{a}_j\right), \forall i\in\mathcal{N}.\end{array}
\end{eqnarray}
Hence, without intervention, the mutual minmax profile may not be a Nash equilibrium. However, with intervention, the mutual minmax profile
$(\bar{a}_0,\bar{a}_1,\ldots,\bar{a}_N)$ is always a Nash equilibrium, as long as the maximum rate $\bar{a}_0$ of the intervention device is high
enough to yield
\begin{eqnarray}
\begin{array}{c}\mu-\sum_{j\neq i} \bar{a}_j - \bar{a}_0 \leq 0,~\forall i\in\mathcal{N}.\end{array}
\end{eqnarray}

To see why we prefer a mutual minmax profile that is an NE, we study the conditions under which a strategy, described by the automaton in this
subsection, is an SPE. For simplicity, consider a special case where $T=\infty$ and
$(\tilde{a}_0^\tau,\mathbf{\tilde{a}}^\tau)=(0,\mathbf{\tilde{a}})~\forall \tau$ in the automaton. When the mutual minmax profile is not an NE, the
conditions on the discount factor $\delta$ and the length of the punishment phase $L$ can be derived from the proof of
Proposition~\ref{proposition:FolkTheorem_Minmax_PureAction} as
\begin{eqnarray}\label{eqn:FlowControl_FolkTheorem_delta_L_1}
\delta+\dots+\delta^L \geq \frac{(\tilde{a}_i^*)^{\beta_i}\left(\mu-\tilde{a}_i^*-\sum_{j\neq i} \tilde{a}_j\right) -
\tilde{a}_i^{\beta_i}\left(\mu-\sum_{j=1}^N \tilde{a}_j\right)}{\tilde{a}_i^{\beta_i}\left(\mu-\sum_{j=1}^N \tilde{a}_j\right) -
\bar{a}_i^{\beta_i}\cdot\max\left\{0,\mu-\bar{a}_0-\sum_{j=1}^N \bar{a}_j\right\}},~\forall i,
\end{eqnarray}
where $\tilde{a}_i^*=\min\left\{\bar{a}_i,\frac{\beta_i}{1+\beta_i}\left(\mu-\sum_{j\neq i} \tilde{a}_j\right)\right\}$, and
\begin{eqnarray}\label{eqn:FlowControl_FolkTheorem_delta_L_2}
\delta^L \geq \frac{(\bar{a}_i^*)^{\beta_i}\cdot\max\left\{0,\mu-\bar{a}_0-\bar{a}_i^*-\sum_{j\neq i} \bar{a}_j\right\} -
\bar{a}_i^{\beta_i}\cdot\max\left\{0,\mu-\bar{a}_0-\sum_{j=1}^N \bar{a}_j\right\}}{\tilde{a}_i^{\beta_i}\left(\mu-\sum_{j=1}^N \tilde{a}_j\right) -
\bar{a}_i^{\beta_i}\cdot\max\left\{0,\mu-\bar{a}_0-\sum_{j=1}^N \bar{a}_j\right\}},~\forall i,
\end{eqnarray}
where $\bar{a}_i^*=\min\left\{\bar{a}_i,\frac{\beta_i}{1+\beta_i}\left(\mu-\bar{a}_0-\sum_{j\neq i} \bar{a}_j\right)\right\}$. For the case without
intervention, we let $\bar{a}_0=0$ in the above inequalities to get the corresponding conditions on $\delta$ and $L$. When the mutual minmax profile
is an NE, only the first inequality needs to be satisfied.

Fig.~\ref{fig:MinDiscountFactor_PunishmentLength_FlowControl_color} shows the minimum discount factor $\delta$ required for the strategy to be an
SPE, under different punishment lengths $L$ and maximum intervention flow rates $\bar{a}_0$. In this example (system parameters shown in the caption
of Fig.~\ref{fig:MinDiscountFactor_PunishmentLength_FlowControl_color}), when the maximum intervention flow rate is $2.5$ bits/s, the mutual minmax
profile is an NE. In this case, we do not need to provide the promise of coming back from the punishment phase. In other words, we allow $L=\infty$.
In this way, we can achieve the minimum discount factor possible. Without intervention or when the maximum intervention rate is $0.5$ bits/s or $1.0$
bits/s, the mutual minmax profile is not an NE. In this case, the minimum discount factor increases when the punishment length increases beyond a
threshold. This is because the users need to be more patient to carry out longer punishment, reflected by
\eqref{eqn:FlowControl_FolkTheorem_delta_L_2}. But note that under the same punishment lengths, the minimum discount factor with the maximum
intervention flow rate $0.5$ bits/s is still smaller than that without intervention. This example indicates that the punishment length should be
carefully designed when the mutual minmax profile is not an NE.
\end{example}

%\vspace{-.5cm}
\subsection{Games With Player-specific Punishments}
Now we study the folk theorem for repeated games with player-specific punishments. The definition of player-specific punishment
\cite[Definition~3.4.1]{MailathSamuelson} can be extended to the case with intervention as follows.
\begin{definition}[Player-specific punishments with intervention]
A payoff $\mathbf{v}\in\mathscr{V}_w^{\dag p}$ allows player-specific punishment if there exists a collection of payoff profiles
$\{\mathbf{v}^i\}_{i=1}^N$, $\mathbf{v}^i\in\mathscr{V}_w^{\dag p}$, such that for all $i$, $v_i>v_i^i$, and for all $j\neq i$, $v_i^j>v_i^i$. The
collection of payoff profiles $\{\mathbf{v}^i\}_{i=1}^N$ is a player-specific punishment for $\mathbf{v}$.
\end{definition}

\begin{proposition}[Folk Theorem With Player-specific Punishments]\label{proposition:FolkTheorem_PlayerSpecific_PureAction}
Suppose that $\mathbf{v}\in\mathscr{V}_w^{\dag p}$ allows player-specific punishments in $\mathscr{V}_w^{\dag p}$. There exists
$\underline{\delta}<1$ such that for all $\delta\in(\underline{\delta},1)$, there exists a subgame perfect equilibrium with payoffs $\mathbf{v}$, of
the repeated game with intervention.
\end{proposition}
\begin{IEEEproof}
See \cite[Appendix~B]{XiaoParkMihaela_Appendix}.
\end{IEEEproof}

\subsubsection{Structure of the equilibrium strategy}
Suppose we want to achieve a SPE payoff $\mathbf{v}$. Again, we denote the equilibrium outcome path by
$\{\tilde{a}_0^\tau,\mathbf{\tilde{a}}^\tau\}_{\tau=0}^{T-1}$ for some integer $T>0$. In the equilibrium strategy, the intervention device and the
users follow the sequence $\{\tilde{a}_0^\tau,\mathbf{\tilde{a}}^\tau\}_{\tau=0}^{T-1}$, or repeat this sequence forever when $T$ is finite. When a
deviation from user $i$ happens, the sequence $\{\hat{a}_0^\ell(i),\mathbf{\hat{a}}^\ell(i)\}_{\ell=0}^{L_i-1}$ is played, generating $\mathbf{v}^i$
in the player-specific punishment. The automaton of this equilibrium strategy is similar to that with mutual minmax profiles, with the only
difference being the specific punishments for different users in this case. Due to space limit, we omit the detailed description of the automaton.

\subsubsection{Discussion and examples}
\begin{example}
We modify the flow control game in Example~2 with another form of intervention, to illustrate that intervention can simplify the player-specific
punishment. We consider an intervention device that inspects the packets at the output of the server. It can identify the sources of the packets and
drop packets with certain probabilities. The intervention device's action is denoted by a vector $\mathbf{a}_0=(a_{0,1},\ldots,a_{0,N})$, with
$a_{0,i}\in[0,1]$ being the probability of dropping user $i$'s packets. Then the stage-game payoff of user $i$ with a fixed $\mathbf{a}_0$ is
\begin{eqnarray}
\begin{array}{c}u_i(\mathbf{a}_0,\mathbf{a}) = ((1-a_{0,i})a_i)^{\beta_i}\cdot\left(\mu-\sum_{j=1}^N a_j \right).\end{array}
\end{eqnarray}

Such an intervention device can carry out player-specific punishments all by itself. If a unilateral deviation occurs, the intervention device will
drop the packets of the deviating user with probability $1$, while it will not drop the other users' packets.
\end{example}

\section{Detailed Analysis of The Protocol Design Problem}\label{sec:DesignProblem}
In Section~\ref{sec:EquilibriumPayoffSet}, we have characterized the set of equilibrium payoffs when the discount factor is sufficiently close to
$1$. With the set of equilibrium payoffs and minimum payoff guarantees, we can obtain the optimal equilibrium payoff $(U_1^*,\ldots,U_N^*)$ that
maximizes the welfare function while satisfying minimum payoff guarantees. In addition, we have obtained the structure of the equilibrium strategy
$(\sigma_0,\sigma)$ that can achieve any equilibrium payoff when the discount factor is sufficiently close to $1$.

In this section, we provide detailed analysis of the protocol design problem under the practical condition that the discount factor is bounded away
from $1$. Specifically, given the optimal equilibrium payoff $(U_1^*,\ldots,U_N^*)$, we derive the minimum discount factor under which the optimal
equilibrium payoff can be achieved by an equilibrium strategy of the structures described in Section~\ref{sec:EquilibriumPayoffSet}, and we construct
the corresponding equilibrium strategy. For analytical tractability, the analysis is carried out for a special class of games to be specified later.
The power control and flow control games in Example~2 and Example~3 are special cases of this class of games.

Note that the derived minimum discount factor is a function of the minimum payoff guarantees and the intervention capability. Thus, we obtain the
trade-off among the discount factor, the minimum payoff guarantees, and the intervention capability. With this trade-off, we can determine if we can
achieve the optimal equilibrium payoff under a given discount factor, and if not, how to change the minimum payoff guarantees such that the optimal
equilibrium payoff can be achieved. Moreover, we can determine the intervention capability required to achieve the optimal equilibrium payoff under a
given (proper) discount factor.

%\vspace{-.5cm}
\subsection{A Special Class of Games}
In the rest of the paper, we consider a special class of games that satisfy two assumptions.
\begin{assumption}
With intervention, the stage game has a pure-action mutual minmax profile $(\hat{a}_0,\mathbf{\hat{a}})$ (see
Definition~\ref{definition:MutualMinmaxProfile}), which is also a Nash equilibrium of the stage game.
\end{assumption}

\begin{remark}
It is very common to have a pure-action mutual minmax profile as a Nash equilibrium in resource allocation games, such as the power control, and flow
control games in Section~\ref{sec:EquilibriumPayoffSet}. Specifically, if all the users share a common resource, i.e., $k_i=1$ for all $i$, and they
consume as much resources as possible, each one of them will be minmaxed. In the power control game in Example~2, the mutual minmax profile is the
NE. In the flow control, although the mutual minmax profile may not be an NE without intervention, using intervention can make it an NE, as we have
shown in Section~\ref{sec:EquilibriumPayoffSet}.
\end{remark}

\begin{assumption}
For each user $i$, there exists an action profile $\mathbf{\check{a}}^i$ such that
\begin{eqnarray}
u_i(\underline{a}_0,\mathbf{\check{a}}^i) = \max_{\mathbf{a}} u_i(\underline{a}_0,\mathbf{a}) \triangleq \bar{v}_i,~\mathrm{and}~u_j(\underline{a}_0,
\mathbf{\check{a}}^i)=0,~\forall j\neq i,
\end{eqnarray}
and that the set of feasible payoffs is $\mathscr{V}_w^{\dag} = \mathrm{co}\left\{(0,\ldots,0),
\mathbf{u}(\underline{a}_0,\mathbf{\check{a}}^1),\ldots,\mathbf{u}(\underline{a}_0,\mathbf{\check{a}}^N)\right\}$.
\end{assumption}

\begin{remark}
This assumption ensures the existence of the action profile $\mathbf{\check{a}}^i$, where user $i$ takes the most advantage of the resources while
the other users receive no benefit. The assumption that $u_j(\underline{a}_0,\mathbf{\check{a}})=0, \forall j\neq i$ is natural in resource
allocation problems, because the other users should consume no resources in order to maximize a specific user's utility, and a user's utility should
be zero when it consumes no resource. In addition, due to significant interference among the users, the set of feasible payoffs is the convex hull of
the $N+1$ points. Assumption~2 holds true for the power control and flow control games in Section~\ref{sec:EquilibriumPayoffSet}. Note that we can
generalize our results to the case where $u_j(\underline{a}_0,\mathbf{\check{a}})>0, \forall j\neq i$, but we will not do that for notational
simplicity.
\end{remark}

With Assumption~2, we can write the Pareto boundary of the feasible set of the protocol design problem explicitly as follows
\begin{eqnarray}\label{eqn:TargetPayoffSet}
\begin{array}{c}\mathscr{P}_{\mathbf{\gamma}} = \left\{\mathbf{v}\in\mathscr{V}_w^{\dag}:\sum_{i=1}^N (v_i/\bar{v}_i)=1,~v_i\geq\gamma_i,~\forall
i\in\mathcal{N}\right\},\end{array}
\end{eqnarray}
To ensure that the Pareto boundary $\mathscr{P}_{\mathbf{\gamma}}$ is nonempty, not singleton, and composed of individually rational payoffs, we
impose the following constraints on the minimum payoff guarantees $\gamma$
\begin{eqnarray}\label{eqn:ConstraintOnMu}
\begin{array}{c}\sum_{i=1}^N (\gamma_i/\bar{v}_i)<1,~\mathrm{and}~\underline{v}_i^w<\gamma_i\leq\bar{v}_i,~\forall i.\end{array}
\end{eqnarray}

%\vspace{-.5cm}
\subsection{Trade-off Among Discount Factor, Minimum Payoff Guarantees, and Intervention Capability}
The optimal equilibrium payoff of the protocol design problem, also referred to as the target payoff $\mathbf{v}^\star$, must lie in the Pareto
boundary $\mathscr{P}_{\gamma}$. Given the target payoff $\mathbf{v}^\star\in\mathscr{P}_{\gamma}$, we determine the minimum discount factor, under
which $\mathbf{v}^\star$ can be achieved as an SPE payoff by an equilibrium strategy of the structure specified in Section~III-A.

In general, it is very difficult to find the minimum discount factor under which a payoff is achieved as an SPE payoff; see \cite{MailathObara} for a
discussion of this topic in repeated prisoners' dilemma. Here, we provide an upper bound $\bar{\delta}(\mathbf{v}^\star)$ of the minimum discount
factor $\delta^\star(\mathbf{v}^\star)$ to achieve the target payoff $\mathbf{v}^\star$ as an SPE payoff. To clearly state the result, we define the
maximum stage-game payoff that user $j$ can get by deviating from any profile in $\left\{(\underline{a}_0,\mathbf{\check{a}}^i)\right\}_{i=1}^N$ as
\begin{eqnarray}
w_j = \max_{i\neq j} \max_{a_j} u_j(\underline{a}_0,a_j,\mathbf{\check{a}}_{-j}^i).
\end{eqnarray}
\begin{theorem}\label{theorem:MinimumDiscountFactor_EquilibriumPayoffSet}
With intervention, the minimum discount factor $\delta^\star(\mathbf{v}^\star)$ to achieve the target payoff $\mathbf{v}^\star$ as an SPE payoff is
upper bounded by
\begin{eqnarray}\label{eqn:MinimumDiscountFactor_UpperBound}
\bar{\delta}(\mathbf{v}^\star) = \max\left\{\max_{j\neq i}
\frac{w_{j}-v_j^\star}{w_{j}-\underline{v}_j^w},~\frac{2(N-1)}{(N-T)+\sqrt{(N-T)^2+4(T-S)(N-1)}}\right\},
\end{eqnarray}
where $T=\sum_{i=1}^N (w_i/\bar{v}_i)$, and $S=\sum_{i=1}^N (\underline{v}_i^w/\bar{v}_i)$.
\end{theorem}
\begin{IEEEproof}
See \cite[Appendix~C]{XiaoParkMihaela_Appendix}.
\end{IEEEproof}

\begin{remark}
First, $\bar{\delta}(\mathbf{v}^\star)$ and the minimum payoff guarantees $\gamma$ is linked through the target payoff $\mathbf{v}^\star$. Different
minimum payoff guarantees result in different target payoffs, which yield different $\bar{\delta}(\mathbf{v}^\star)$. Second,
$\bar{\delta}(\mathbf{v}^\star)$ is an increasing function of the minmax payoff $\left\{\underline{v}_i^w\right\}_{i=1}^N$, which is determined by
the intervention capability $A_0$. Since $\underline{v}_i^w<\underline{v}_i^o$ for all $i$, applying intervention to repeated games lowers the
minimum requirement on the discount factor, thus supports less patient users in a more dynamic network.
\end{remark}

\subsection{The Equilibrium Strategy}
With Assumption~1, we can use the equilibrium strategy in which the punishment for deviation is playing the mutual minmax profile forever. The
automon of this equilibrium strategy is a simplification of the one shown in Fig.~\ref{fig:Automata_RepeatedGames_NotNE}. The set of states is
$\mathscr{W}=\mathscr{W}_e\cup\{w_p\}$ with the initial state $w_e(0)$, where $\mathscr{W}_e=\left\{w_e(\tau):0\leq \tau\leq T-1\right\}$ is the set
of states in the equilibrium outcome path, and $w_p$ is the state in the punishment phase. The action profile at each state is specified by the
output function
\begin{eqnarray}
f(w)=\left\{\begin{array}{ll} (\tilde{a}_0^\tau,\mathbf{\tilde{a}}^\tau), & \mathrm{if}~w=w_e(\tau) \\ (\hat{a}_0,\mathbf{\hat{a}}), &
\mathrm{if}~w=w_p\end{array}\right..
\end{eqnarray}
The state transition is specified by $\lambda(w_p,(a_0,\mathbf{a}))=w_p~\forall (a_0,\mathbf{a})$, and
\begin{eqnarray}
\lambda(w_e(\tau),(a_0,\mathbf{a}))=\left\{\begin{array}{ll} w_e(\tau+1 \mod T), & \mathrm{if}~(a_0,\mathbf{a})=(\tilde{a}_0^\tau,\mathbf{\tilde{a}}^\tau) \\
w_p, & \mathrm{otherwise} \end{array}\right..
\end{eqnarray}

With Assumption~2, it is sufficient to choose every action profile $(\tilde{a}_0^\tau,\mathbf{\tilde{a}}^\tau)$ in the equilibrium outcome path from
the set $\left\{(\underline{a}_0,\mathbf{\check{a}}^i)\right\}_{i=1}^N$.

Now the only unknown is the users' action profiles in the equilibrium outcome path, i.e., $\{\mathbf{\tilde{a}}^\tau\}_{\tau=0}^{T-1}$. There are two
requirements for $\{\mathbf{\tilde{a}}^\tau\}_{\tau=0}^{T-1}$. First, it results in an equilibrium outcome path whose discounted average payoff is
the target payoff $\mathbf{v}^\star$ under a given discount factor. Second, it results in a strategy, described in the above automaton, that is an
SPE under the given discount factor. Theorem~\ref{theorem:EquilibriumStrategy} shows how to design $\{\mathbf{\tilde{a}}^\tau\}_{\tau=0}^{T-1}$ for
any target payoff $\mathbf{v}^\star\in\mathscr{P}_{\gamma}$, under any discount factor $\delta\geq\bar{\delta}(\mathbf{v}^\star)$.

\begin{theorem}\label{theorem:EquilibriumStrategy}
For any target payoff $\mathbf{v}^\star\in\mathscr{P}_{\gamma}$, and any discount factor $\delta\geq\bar{\delta}(\mathbf{v}^\star)$, the users'
action profiles in the equilibrium outcome path, $\{\mathbf{\tilde{a}}^\tau\}_{\tau=0}^{T-1}$, can be generated by the algorithm in
Table~\ref{table:EquilibriumStrategy}.
\end{theorem}
\begin{IEEEproof}
See \cite[Appendix~D]{XiaoParkMihaela_Appendix}.
\end{IEEEproof}

\begin{remark}
Note that any action profile $\mathbf{\tilde{a}}(\tau)$ in the sequence is from the set of $\{\mathbf{\tilde{a}}^i\}_{i=1}^N$. In other words, only
one user takes nonzero action in each period. This greatly simplifies the monitoring burden of the intervention device and the users. Actually, at
each period, the inactive users, who take zero actions, do not need to monitor, because the active user, who takes nonzero action at that period, can
sense the interference and report to the intervention device the detection of deviation. Then the intervention device can broadcast the detection of
deviation to trigger the punishment. In addition, no inactive user can gain from sending false report to the intervention device, because the
intervention device only trusts the report from the active user.
\end{remark}

\section{Simulation Results}\label{sec:Simulation}
In this section, we consider the flow control game in Example~2 to illustrate the performance gain of using intervention in repeated games and the
trade-off among the discount factor, the minimum payoff guarantees, and the maximum intervention flow rate.

\subsection{Performance Gain}
First, we compare the performance when the protocol designer solves the protocol design problem (\ref{eqn:ProtocolDesignProblem}) using four
different schemes, namely greedy algorithms \cite{BharathKumarJaffe}--\cite{ZhangDouligeris_TCOM}, one-shot games with incentive schemes
\cite{ParkMihaela_JSAC}\cite{GaiKrishnamachari_Infocom}\cite{YiMihaela_TCOM}, repeated games without intervention, and repeated games with
intervention. The two example welfare functions we use are the sum payoff $\sum_{i\in\mathcal{N}} U_i$ and the absolute fairness
$\min_{i\in\mathcal{N}} U_i$. Greedy algorithms achieve the NE, which may not satisfy the minimum payoff guarantees. For one-shot games with
incentive schemes, we assume that the entire Pareto boundary of the set of pure-action payoffs can be achieved as NE by using appropriate incentive
schemes, in order to get the best performance achievable by this scheme.

\subsubsection{Impact of the number of users}
We compare how the performance scales with the number of users under the four schemes. For simplicity, we study the symmetric case, where all the
users have the same throughput-delay trade-off $\beta=3$ and the same maximum flow rate normalized to $1$ bits/s. We consider two scenarios: first,
the server's capacity (service rate) increases linearly with the number of users, i.e. $\mu=N$ bits/s, and second, the server's capacity is limited
at $10$ bits/s, i.e. $\mu=\min\{N,10\}$ bits/s. The maximum intervention flow rate is $\bar{a}_0=\max\{\mu-(N-1),0\}$ bits/s to ensure the mutual
minmax profile is an NE. We set the minimum payoff as 10\% of the maximum stage-game payoff $\bar{v}_i$ subject to the constraints
(\ref{eqn:ConstraintOnMu}), namely $\gamma_i=\min\{0.1\cdot\bar{v}_i,~\mu/N,~\underline{v}_i^w\}$ bits/s.

\begin{figure}
\centering
\includegraphics[width =4.0in]{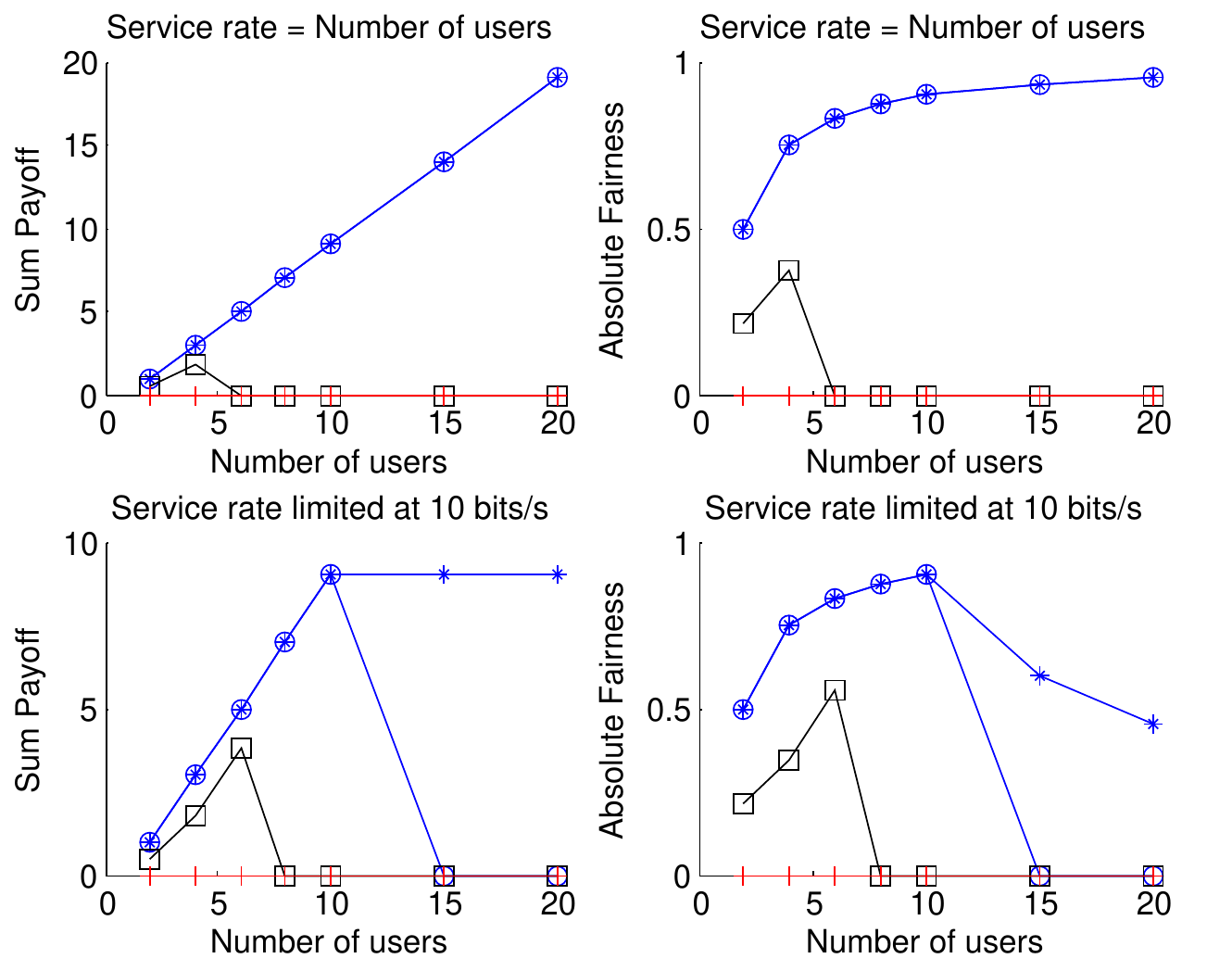}
\caption{Performance comparison among the four schemes with the number of users increasing. Blue lines with asterisks: repeated games with
intervention, blue lines with circles: repeated games without intervention, black lines with squares: one-shot games with incentive schemes, red
lines with crosses: Nash equilibrium of the one-shot game.} \label{fig:PerformanceComparison_ScaleWithUserNumber_color}
\end{figure}

Fig.~\ref{fig:PerformanceComparison_ScaleWithUserNumber_color} shows the sum payoff and the fairness achieved by different schemes. When the capacity
increases linearly with the number of users $N$, both the sum payoff and the fairness increase with $N$ by using repeated games. In contrast, when
using one-shot games with incentive schemes, the sum payoff and fairness increase initially when the number of users is small, and decrease when
$N>4$. A sum payoff or fairness of the value $0$ means that the minimum payoff cannot be guaranteed. This happens in one shot games with incentive
schemes when $N>5$. The NE payoff does not satisfy the minimum payoff guarantee with any number of users.

When the capacity is limited at $10$ bits/s, for repeated games with intervention, the sum payoff reaches the bottleneck of $10$, and due to
congestion, the fairness decreases when $N>10$. For repeated games without intervention, the sum payoff and the fairness decrease more rapidly, and
the minimum payoff guarantee cannot be met when $N\geq 15$. For one-shot games, the trend of the sum payoff and fairness is similar to the case with
increasing capacity.

In conclusion, using repeated games with intervention has a large performance gain over the other three schemes, in terms of both the sum payoff and
the fairness.

\subsubsection{Impact of minimum payoff guarantees}
We compare the performance of the four schemes under different minimum payoff guarantees. The system parameters are the same as those in
Fig.~\ref{fig:MinDiscountFactor_PunishmentLength_FlowControl_color}. The maximum intervention flow rate is $\bar{a}_0=2.5$ bits/s to ensure the
mutual minmax profile is an NE. We set the same minimum payoff guarantee for all the users, namely $\gamma_i=\gamma_j,\forall i,j\in\mathcal{N}$.

Table~\ref{table:PerformanceComparison} shows the sum payoff and the fairness achieved by the four schemes. We write ``N/A'' when the minimum payoff
guarantee cannot be satisfied. For repeated games, we show the minimum discount factors allowed to achieve the optimal performance in the parenthesis
next to the performance metric. An immediate observation is the inefficiency of the NE, as expected.

When using one-shot game model with incentive schemes, the performance loss compared to using repeated games is small when the minimum payoff
guarantee is small ($\gamma_i=1$). However, when the minimum payoff guarantee increases, using one-shot games is far from optimality in terms of both
the sum payoff and the fairness. Note that using one-shot games fails to satisfy the large minimum payoff guarantee ($\gamma_i=14$). In summary,
using repeated games has large performance gain over using one-shot games in most cases, and is necessary when the minimum payoff guarantee is large.

\begin{figure}
\centering
\includegraphics[width =4.0in]{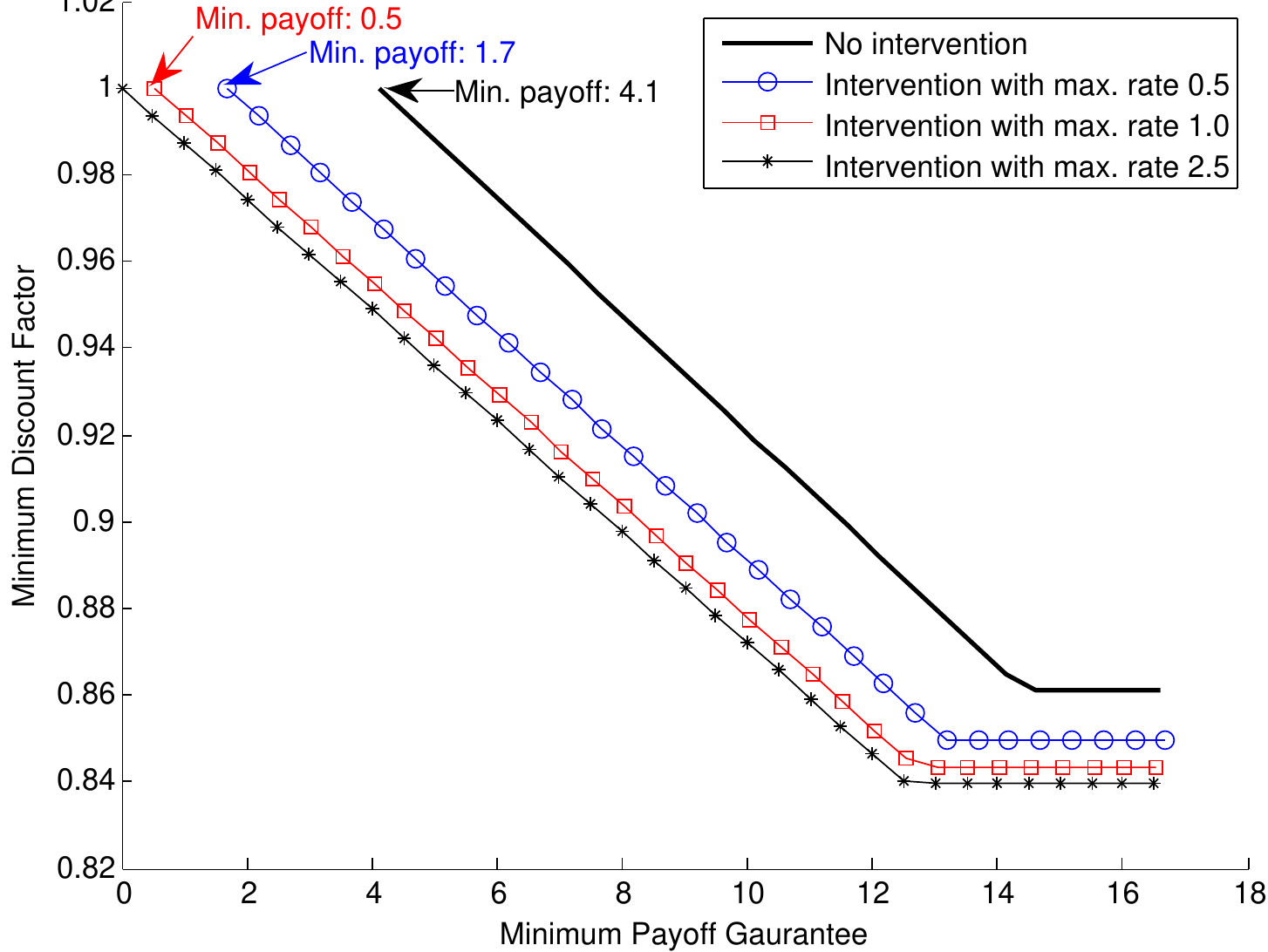}
\caption{The trade-offs between the minimum discount factor and the minimum payoff guarantee under different maximum intervention flow rates. At the
beginning of each curve, we mark the smallest minimum payoff guarantees we can impose, which indicates the largest feasible set of the protocol
design problem with each maximum intervention flow rate.} \label{fig:Tradeoff_DiscountfactorQoS_FlowControl4user_color}
\end{figure}

Under the specific parameters in this simulation, if we allow the discount factor to be sufficiently close to $1$, using intervention in repeated
games has little performance gain over repeated games without intervention, especially when the minimum payoff guarantee is large. This is because
the minmax payoff without intervention is already small, such that its is Pareto dominated by some large minimum payoff guarantees. In this case, the
advantage of using intervention in repeated games is that it allows smaller discount factors to achieve the same or better performance, compared to
using repeated games without intervention. This is also confirmed by Fig.~\ref{fig:Tradeoff_DiscountfactorQoS_FlowControl4user_color}, which shows
the minimum discount factor allowed to achieve the target payoff that maximizes the sum payoff, under different minimum payoffs $\gamma$ and
different maximum intervention flow rates.

%\vspace{-.5cm}
\subsection{Trade-off among $\delta$, $\gamma$, and $\bar{a}_0$}
\begin{figure}
\centering
\includegraphics[width =4.0in]{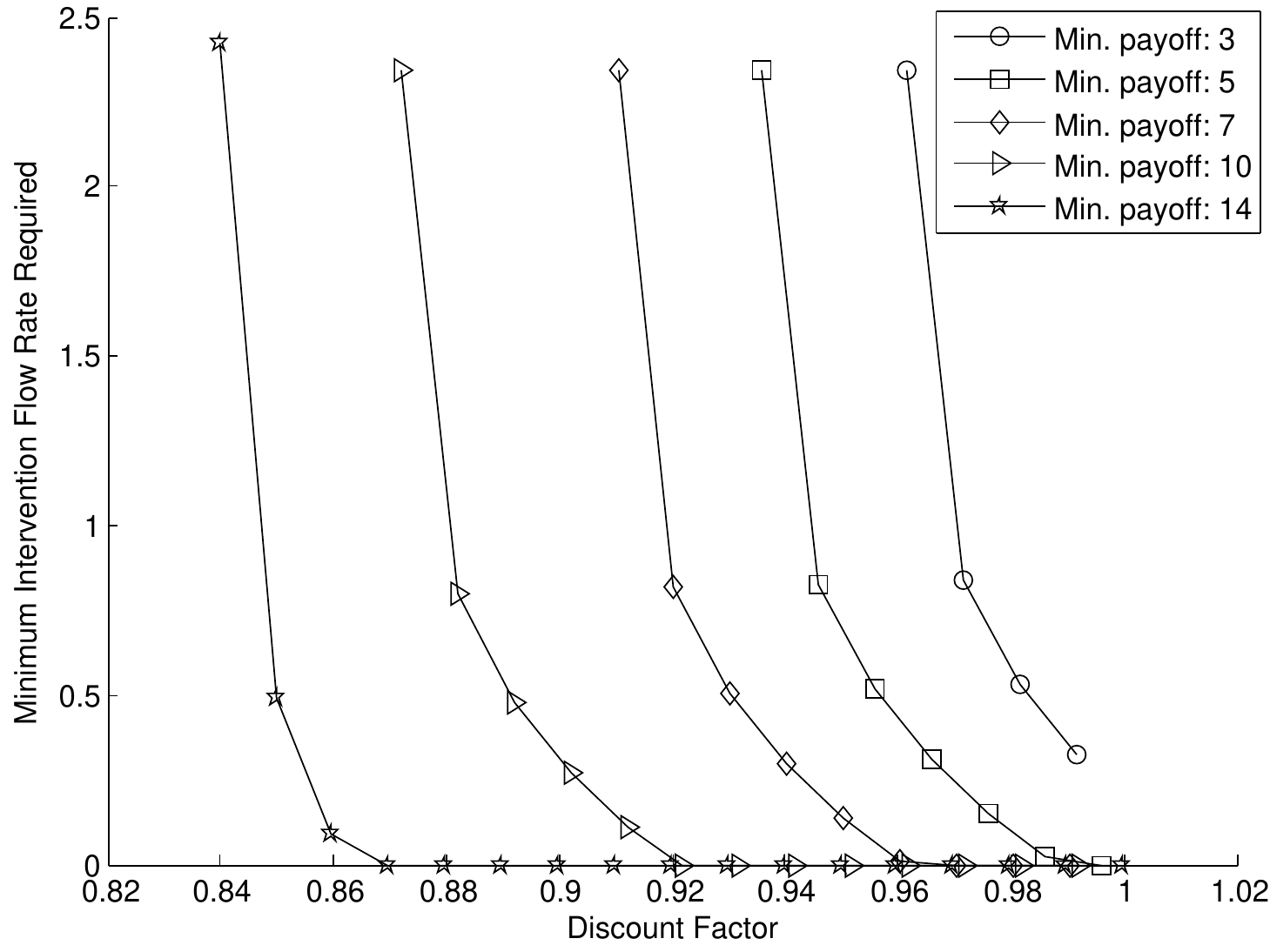}
\caption{The trade-offs between the required maximum intervention flow rate and the discount factor under different minimum payoff guarantees.}
\label{fig:Tradeoff_InterventionrateDiscountfactor_FlowControl4user}
\end{figure}

\begin{figure}
\centering
\includegraphics[width =4.0in]{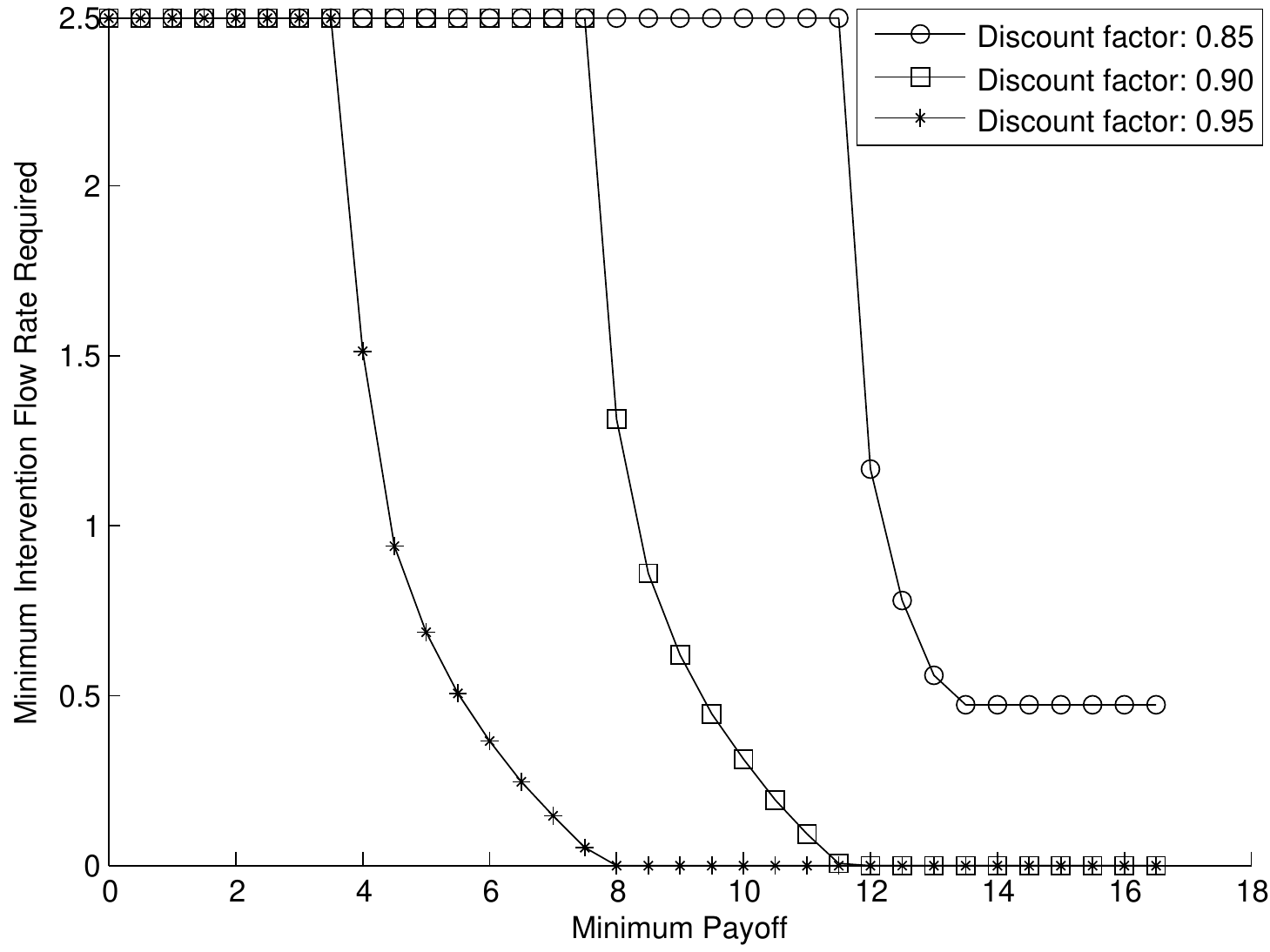}
\caption{The trade-offs between the required maximum intervention flow rate and the minimum payoff guarantee under different discount factors.}
\label{fig:Tradeoff_InterventionrateQoS_FlowControl4user}
\end{figure}

Consider the same system as that in Fig.~\ref{fig:MinDiscountFactor_PunishmentLength_FlowControl_color} and
Fig.~\ref{fig:Tradeoff_DiscountfactorQoS_FlowControl4user_color}. Suppose that the target payoff is the one that maximizes the sum payoff. In
Fig.~\ref{fig:Tradeoff_InterventionrateDiscountfactor_FlowControl4user}, we plot the trade-off between the required maximum intervention flow rate
and the discount factor under different minimum payoff guarantees. In Fig.~\ref{fig:Tradeoff_InterventionrateQoS_FlowControl4user}, we plot the
trade-off between the required maximum intervention rate and the minimum payoff guarantees under different discount factors. The protocol designer
can use these trade-off curves as guidelines for determining the maximum intervention flow rate required under different discount factors and minimum
payoff guarantees.

\section{Conclusion}\label{sec:Conclusion}
In this paper, we proposed a repeated game model generalized by intervention, which can be applied to a large variety of communication systems. We
use repeated games to achieve the equilibrium payoffs that Pareto dominate some payoffs on the Pareto boundary of the nonconvex set of feasible
payoffs in one-shot games. In addition, we combine conventional repeated games with intervention, and show that intervention enlarges the set of
equilibrium payoffs when the discount factor is sufficiently close to $1$. Then we consider the  protocol design problem of maximizing the welfare
function subject to minimum payoff guarantees. We derive the trade-off between the discount factor, the minimum payoff guarantees, and the
intervention capability, and construct equilibrium strategies to achieve any target payoff as an SPE payoff. Simulation results show the great
performance gain, in terms of sum payoff and absolute fairness, by using intervention in repeated games.

% Appendix

\appendices
%\section{Overview of Automata}\label{appendix:Automata}

\section{Proof of Proposition~1}\label{appendix:FolkTheorem_Minmax}
\begin{figure}
\centering
\includegraphics[width =6.0in]{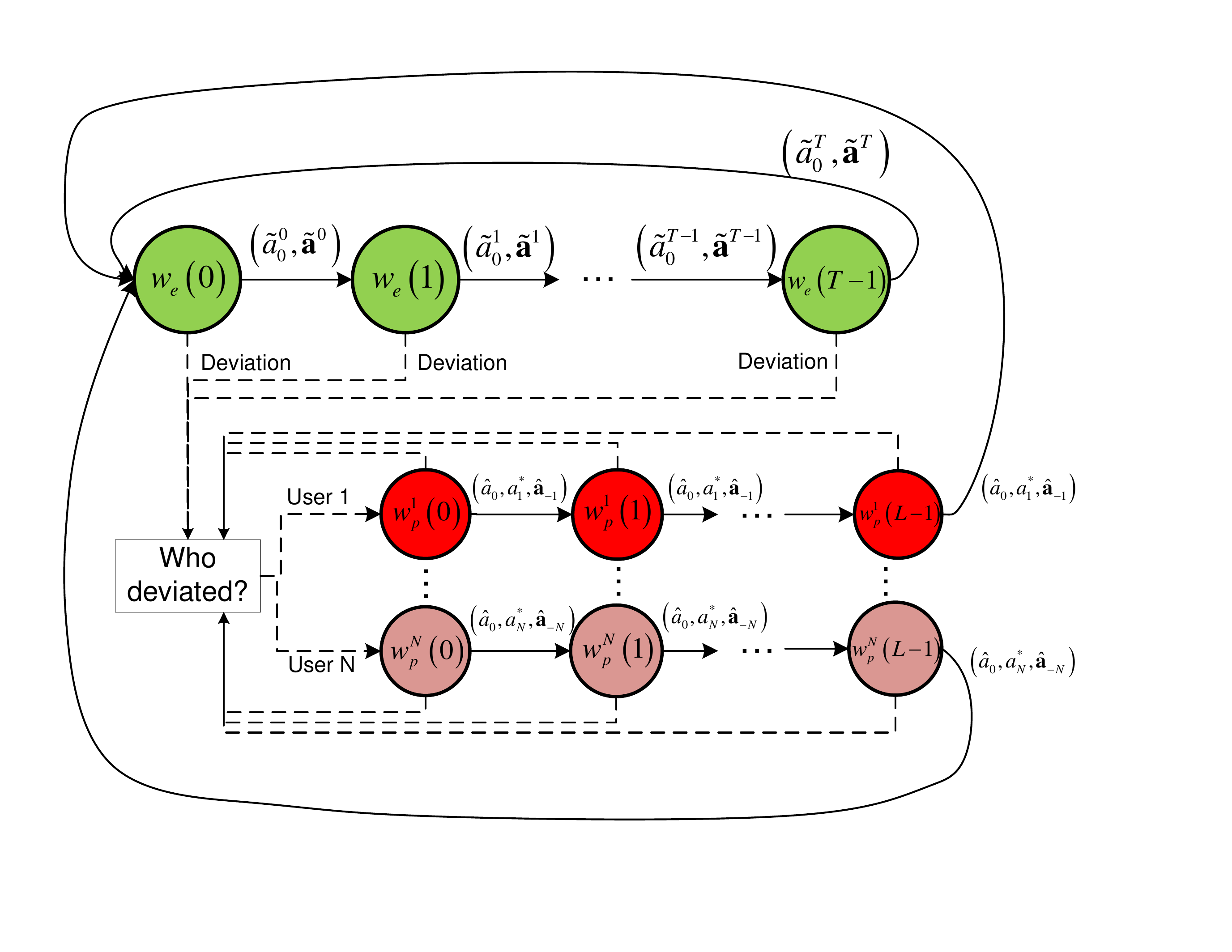}
\caption{The automaton of the equilibrium strategy of the game with a mutual minmax profile $(\hat{a}_0,\mathbf{\hat{a}})$. Circles are states, where
$\left\{w_e(\tau)\right\}_{\tau=0}^{T-1}$ is the set of states in the equilibrium outcome path, and $\left\{w_p^i(\ell)\right\}_{\ell=0}^{L-1}$ is
the set of states in the punishment phase for user $i$. The initial state is $w_e(0)$. Solid arrows are the prescribed state transitions labeled by
the action profiles leading to the transitions. Dashed arrows are the state transitions when deviation happens. $a_i^*$ is user $i$'s best response
to $\hat{a}_0$ and $\mathbf{\hat{a}_{-i}}$.} \label{fig:Automata_RepeatedGames_NotNE}
\end{figure}

We prove the minmax folk theorem by constructing an equilibrium strategy profile presented in the automaton in
Fig.~\ref{fig:Automata_RepeatedGames_NotNE}. Assume that the game has a mutual minmax profile $(\hat{a}_0,\mathbf{\hat{a}})$. First, we prove that
any pure-action payoff profile $\mathbf{u}(\tilde{a}_0,\mathbf{\tilde{a}})$ that Pareto dominates the minmax payoff profile
$\mathbf{u}(\hat{a}_0,\mathbf{\hat{a}})$ can be achieved as an SPE payoff. Then, we prove that any payoff profile $\mathbf{v}\in\mathscr{V}_w^{\dag
p}$, which may not be achieved by a pure-action profile, can be achieved as an SPE payoff.

For the reader's convenience, we rewrite the formal description of the automaton here:
\begin{itemize}
\item The set of states is $\mathscr{W}=\mathscr{W}_e\cup\mathscr{W}_p^1\cup\cdots\cup\mathscr{W}_p^N$, where $\mathscr{W}_e=\left\{w_e(\tau):0\leq \tau\leq T-1\right\}$ is the set of states
in the equilibrium outcome path, and $\mathscr{W}_p^i=\left\{w_p^i(\ell):0\leq \ell\leq L-1\right\}$ is the set of states in the punishment phase for
user $i$. The initial state is $w^0=w_e(0)$.
\item The action profile at each state is specified by the following output function
\begin{eqnarray}
f(w)=\left\{\begin{array}{ll} (\tilde{a}_0^\tau,\mathbf{\tilde{a}}^\tau), & \mathrm{if}~w=w_e(\tau) \\ (\hat{a}_0,a_i^*,\mathbf{\hat{a}}_{-i}), &
\mathrm{if}~w\in\mathscr{W}_p^i\end{array}\right.,
\end{eqnarray}
where $a_i^*$ is user $i$'s best response to $\hat{a}_0$ and $\mathbf{\hat{a}}_{-i}$.
\item The state transition is specified by the following transition rule
\begin{eqnarray}
\lambda(w_e(\tau),(a_0,\mathbf{a}))=\left\{\begin{array}{ll} w_p^i(0), & \mathrm{if}~a_i\neq\tilde{a}_i^\tau~\mathrm{and}~\mathbf{a}_{-i}=\mathbf{\tilde{a}}_{-i}^\tau \\
w_e(\tau+1 \mod T), & \mathrm{otherwise} \end{array}\right.,
\end{eqnarray}
and
\begin{eqnarray}
\lambda(w_p^i(\ell),(a_0,\mathbf{a}))=\left\{\begin{array}{ll} w_p^i(\ell+1), & \mathrm{if}~\ell<L-1~\mathrm{and}~(a_0,\mathbf{a})=(\hat{a}_0,a_i^*,\mathbf{\hat{a}}_{-i}) \\
w_e(0), & \mathrm{if}~\ell=L~\mathrm{and}~(a_0,\mathbf{a})=(\hat{a}_0,a_i^*,\mathbf{\hat{a}}_{-i}) \\
w_p^i(0), & \mathrm{if}~a_i\neq a_i^*~\mathrm{and}~\mathbf{a}_{-i}=\mathbf{\hat{a}}_{-i}^\tau \\
w_p^j(0), & \mathrm{if}~j\neq i,~a_j\neq \hat{a}_j,~a_i=a_i^*,~\mathrm{and}~a_k=\hat{a}_k~\forall k\neq i,j \end{array}\right..
\end{eqnarray}
\end{itemize}

\subsection{Achieving Pure-action Payoff As an SPE Payoff}
We prove that for any pure-action payoff profile $\mathbf{u}(\tilde{a}_0,\mathbf{\tilde{a}})$ that Pareto dominates the minmax payoff profile
$\mathbf{u}(\hat{a}_0,\mathbf{\hat{a}})$, there exists $\underline{\delta}<1$ and a strategy profile presented in the automaton in
Fig.~\ref{fig:Automata_RepeatedGames_NotNE}, such that for all $\delta\in(\underline{\delta},1)$, the strategy profile is a subgame perfect
equilibrium with payoff $\mathbf{u}(\tilde{a}_0,\mathbf{\tilde{a}})$.

Since the payoff can be achieved by a pure-action profile $(\tilde{a}_0,\mathbf{\tilde{a}})$, the equilibrium outcome path is repeating
$(\tilde{a}_0,\mathbf{\tilde{a}})$ in every period. Hence, the automaton is simplified to have $T=1$ and
$(\tilde{a}_0^0,\mathbf{\tilde{a}}^0)=(\tilde{a}_0,\mathbf{\tilde{a}})$.

Now we calculate the values of all the states in the automaton. For $w_e(0)$, we have
\begin{eqnarray}
\mathbf{V}(w_e(0))=(1-\delta)\cdot\mathbf{u}(\tilde{a}_0,\mathbf{\tilde{a}}) + \delta\cdot \mathbf{V}(w_e(0)) \Rightarrow
\mathbf{V}(w_e(0))=\mathbf{u}(\tilde{a}_0,\mathbf{\tilde{a}}).
\end{eqnarray}
For $w_p^i(\ell),0\leq \ell<L-1$, we have
\begin{eqnarray}
\mathbf{V}(w_p^i(\ell)) &=& (1-\delta) \cdot \mathbf{u}(\hat{a}_0,a_i^*,\mathbf{\hat{a}_{-i}})+\delta \cdot \mathbf{V}(w_p^i(\ell+1)) \\
&=& (1-\delta) \cdot \left[\mathbf{u}(\hat{a}_0,a_i^*,\mathbf{\hat{a}_{-i}})\cdot\frac{1-\delta^{L-\ell-1}}{1-\delta}\right] + \delta^{L-\ell-1}
\cdot \mathbf{V}(w_p^i(L-1))
\end{eqnarray}
Since we can calculate $\mathbf{V}(w_p^i(L-1))$ as
\begin{eqnarray}
\mathbf{V}(w_p^i(L-1)) = (1-\delta)\cdot\mathbf{u}(\hat{a}_0,a_i^*,\mathbf{\hat{a}_{-i}}) + \delta\cdot\mathbf{V}(w_e(0)),
\end{eqnarray}
we have for $\ell=0,\ldots,L-1$,
\begin{eqnarray}
\mathbf{V}(w_p^i(\ell)) = (1-\delta^{L-\ell})\cdot\mathbf{u}(\hat{a}_0,a_i^*,\mathbf{\hat{a}_{-i}}) +
\delta^{L-\ell}\cdot\mathbf{u}(\tilde{a}_0,\mathbf{\tilde{a}}).
\end{eqnarray}

With the values of the states, we can derive the conditions under which the strategy prescribed by this automaton is a subgame perfect equilibrium.
To this end, we need to check that for any state $w$ accessible from $w^0$, $f(w)$ is a Nash equilibrium of the normal-form game described by the
payoff function $g^w:\mathcal{A}\rightarrow\mathbb{R}^N$, where
\begin{eqnarray}
g^w(\mathbf{a})=(1-\delta)\cdot\mathbf{u}(a_0(w),\mathbf{a})+\delta\cdot\mathbf{V}(\lambda(w,a_0(w),\mathbf{a})),
\end{eqnarray}
where $a_0(w)$ is the intervention action prescribed in $f(w)$.

First, we check the incentive compatibility constraints for users to follow the action profile $\mathbf{\tilde{a}}$. This can be done by checking if
$\mathbf{\tilde{a}}$ is a Nash equilibrium of the normal-form game at the state $w=w_e(0)$. The normal-form game at state $w=w_e(0)$ has payoff
function
\begin{eqnarray}
g^w(\mathbf{a})=\left\{\begin{array}{ll} (1-\delta)\cdot\mathbf{u}(\tilde{a}_0,\mathbf{\tilde{a}})+\delta\cdot\mathbf{V}(w_e(0)) =
\mathbf{u}(\tilde{a}_0,\mathbf{\tilde{a}}), & \mathrm{if}~\mathbf{a} = \mathbf{\tilde{a}} \\
(1-\delta)\cdot\mathbf{u}(\tilde{a}_0,\mathbf{a})+\delta\cdot\mathbf{V}(w_p^i(0)), &
\mathrm{if}~a_i\neq\tilde{a}_i~\mathrm{and}~\mathbf{a}_{-i}=\mathbf{\tilde{a}}_{-i}
\end{array}\right..
\end{eqnarray}
The action profile $\mathbf{\tilde{a}}$ is a Nash equilibrium if and only if, for all $i\in\mathcal{N}$ and for all $a_i\in A_i$,
\begin{eqnarray}
u_i(\tilde{a}_0,\mathbf{\tilde{a}}) \geq (1-\delta)\cdot u_i(\tilde{a}_0,a_i,\mathbf{\tilde{a}_{-i}}) + \delta\cdot\left[(1-\delta^{L})\cdot
u_i(\hat{a}_0,\mathbf{\hat{a}}) + \delta^{L}\cdot u_i(\tilde{a}_0,\mathbf{\tilde{a}})\right].
\end{eqnarray}
Define $M\triangleq\max_{i,\mathbf{a}} u_i(\tilde{a}_0^i,\mathbf{a})$. Then it suffices to have, for all $i\in\mathcal{N}$,
\begin{eqnarray}
(1-\delta^{L+1})\cdot u_i(\tilde{a}_0,\mathbf{\tilde{a}}) \geq (1-\delta)\cdot M + \delta\cdot(1-\delta^{L})\cdot u_i(\hat{a}_0,\mathbf{\hat{a}}).
\end{eqnarray}
After rearranging the terms, we have
\begin{eqnarray}\label{eqn:FolkTheorem_Minmax_delta_L_1}
(\delta+\dots+\delta^L)\cdot (u_i(\tilde{a}_0,\mathbf{\tilde{a}})-u_i(\hat{a}_0,\mathbf{\hat{a}}))\geq M - u_i(\tilde{a}_0,\mathbf{\tilde{a}}).
\end{eqnarray}

Second, we check the incentive compatibility constraints for staying in the punishment phase. This is done by checking if $\mathbf{\hat{a}}$ is a
Nash equilibrium of the normal-form games at the states $w_p^i(0),\ldots,w_p^i(L-1)$. Actually, we only need the incentive compatibility constraint
to hold for the game at state $w_p^i(0)$, because the value of state $w_p^i(0)$ is the lowest among all the states in the punishment phase, and user
$i$'s deviation to $a_i$ in any state in the punishment phase results in the same payoff of
\begin{eqnarray}
(1-\delta)\cdot u_i(\hat{a}_0,a_i,\mathbf{\hat{a}}_{-i}) + \delta\cdot V_i(w_p^i(0)).
\end{eqnarray}
Hence, for the punishing action profile $\mathbf{\hat{a}}$ to be incentive compatible, we need for all $i\in\mathcal{N}$ and for all $a_i\in A_i$,
\begin{eqnarray}
V_i(w_p^i(0)) \geq (1-\delta)\cdot u_i(\hat{a}_0,a_i,\mathbf{\hat{a}}_{-i}) + \delta\cdot V_i(w_p^i(0)),
\end{eqnarray}
which gives us for all $i\in\mathcal{N}$ and for all $a_i\in A_i$,
\begin{eqnarray}
V_i(w_p^i(0)) = (1-\delta^{L})\cdot u_i(\hat{a}_0,\mathbf{\hat{a}}) + \delta^{L}\cdot u_i(\tilde{a}_0,\mathbf{\tilde{a}}) \geq
u_i(\hat{a}_0,a_i,\mathbf{\hat{a}}_{-i}).
\end{eqnarray}
Since $u_i(\hat{a}_0,a_i,\mathbf{\hat{a}}_{-i})\leq \underline{v}_i^w$, it suffices to have for all $i\in\mathcal{N}$,
\begin{eqnarray}\label{eqn:FolkTheorem_Minmax_delta_L_2}
(1-\delta^{L})\cdot u_i(\hat{a}_0,\mathbf{\hat{a}}) + \delta^{L}\cdot u_i(\tilde{a}_0,\mathbf{\tilde{a}}) \geq \underline{v}_i^w.
\end{eqnarray}

Now we have two sets of constraints (\ref{eqn:FolkTheorem_Minmax_delta_L_1}) and (\ref{eqn:FolkTheorem_Minmax_delta_L_2}) for the discount factor
$\delta$ and the length of punishment $L$. We choose $L$ such that $L\cdot (u_i(\tilde{a}_0,\mathbf{\tilde{a}})-u_i(\hat{a}_0,\mathbf{\hat{a}})) > M
- u_i(\tilde{a}_0,\mathbf{\tilde{a}})$ for all $i$. Then we can find a discount factor $\underline{\delta}^p<1$ such that
(\ref{eqn:FolkTheorem_Minmax_delta_L_1}) and (\ref{eqn:FolkTheorem_Minmax_delta_L_2}) are satisfied for all $i$.

\subsection{Achieving Any Feasible and Individually Rational Payoff}
We prove that any feasible and individually rational payoff $\mathbf{v}\in\mathscr{V}_w^{\dag p}$ can be achieved as an SPE payoff. The difficulty is
that the payoff starting from a certain period may be too low to prevent users from deviating at that period. This difficulty is resolved by
\cite[Lemma 3.7.2]{MailathSamuelson}, which states that for any $\varepsilon>0$, there exists $\underline{\delta}^\prime<1$ such that for any payoff
$\mathbf{v}\in\mathscr{V}_w^{\dag p}$ and any discount factor $\delta\in(\underline{\delta}^\prime,1)$, there exists a sequence of pure action
profiles that gives discounted average payoff $\mathbf{v}$ and that has continuation payoffs within $\varepsilon$ of $\mathbf{v}$ at any period $t$.
In other words, the payoff starting from any period is approximately the same, as if generated by a pure-action profile. Combining with the results
in the previous subsection, there exists $\underline{\delta}=\max\{\underline{\delta}^p,\underline{\delta}^\prime\}$, such that any discount factor
$\delta\in(\underline{\delta},1)$ can sustain $\mathbf{v}\in\mathscr{V}_w^{\dag p}$ as an SPE payoff.

\section{Proof of Proposition~2}\label{appendix:FolkTheorem_PlayerSpecific}
We prove the folk theorem with player-specific punishments by constructing an equilibrium strategy profile. First, we prove that any pure-action
payoff profile $\mathbf{v}=\mathbf{u}(\tilde{a}_0,\mathbf{\tilde{a}})$ that allows pure-action player-specific punishments
$\{\mathbf{v}^i=\mathbf{u}(\hat{a}_0(i),\mathbf{\hat{a}}(i))\}_{i=1}^N$ can be achieved as an SPE payoff. Then, we prove that any payoff profile
$\mathbf{v}\in\mathscr{V}_w^{\dag p}$ that allows player-specific punishments, which may not be achieved by pure-action profiles, can be achieved as
an SPE payoff.

For the reader's convenience, we write the formal description of the automaton (when using pure-action player-specific punishments) here:
\begin{itemize}
\item The set of states is $\mathscr{W}=\mathscr{W}_e\cup\mathscr{W}_p^1\cup\cdots\cup\mathscr{W}_p^N$, where $\mathscr{W}_e=\left\{w_e(\tau):0\leq \tau\leq T-1\right\}$ is the set of states
in the equilibrium outcome path, and $\mathscr{W}_p^i=\left\{w_p^i(\ell):0\leq \ell\leq L\right\}$ is the set of states in the punishment phase for
user $i$. The initial state is $w^0=w_e(0)$.
\item The action profile at each state is specified by the following output function
\begin{eqnarray}
f(w)=\left\{\begin{array}{ll} (\tilde{a}_0^\tau,\mathbf{\tilde{a}}^\tau), & \mathrm{if}~w=w_e(\tau) \\ (\hat{a}_0^i,a_i^*,\mathbf{\hat{a}}_{-i}^i), &
\mathrm{if}~w= w_p^i(\ell)~\mathrm{and}~\ell<L \\ (\hat{a}_0(i),\mathbf{\hat{a}}(i)), & \mathrm{if}~w= w_p^i(L)\end{array}\right.,
\end{eqnarray}
where $(\hat{a}_0^i,a_i^*,\mathbf{\hat{a}}_{-i}^i)$ is the action profile that minmaxes user $i$, namely
\begin{eqnarray}
\min_{a_0,\mathbf{a}_{-i}} \max_{a_i} u_i(a_0,a_i,\mathbf{a}_{-i})=u_i(\hat{a}_0^i,a_i^*,\mathbf{\hat{a}}_{-i}^i)\triangleq \underline{v}_i^w.
\end{eqnarray}
\item The state transition is specified by the following transition rule
\begin{eqnarray}
\lambda(w_e(\tau),(a_0,\mathbf{a}))=\left\{\begin{array}{ll} w_p^i(0), & \mathrm{if}~a_i\neq\tilde{a}_i^\tau~\mathrm{and}~\mathbf{a}_{-i}=\mathbf{\tilde{a}}_{-i}^\tau \\
w_e(\tau+1 \mod T), & \mathrm{otherwise} \end{array}\right.,
\end{eqnarray}
and
\begin{eqnarray}
\lambda(w_p^i(\ell),(a_0,\mathbf{a}))=\left\{\begin{array}{ll} w_p^i(\ell+1), & \mathrm{if}~\ell<L~\mathrm{and}~(a_0,\mathbf{a})=(\hat{a}_0^i,a_i^*,\mathbf{\hat{a}}_{-i}^i) \\
w_p^i(L), & \mathrm{if}~\ell=L~\mathrm{and}~(a_0,\mathbf{a})=(\hat{a}_0(i),\mathbf{\hat{a}}(i)) \\
w_p^i(0), & \mathrm{if}~a_i\neq a_i^*,~\mathbf{a}_{-i}=\mathbf{\hat{a}}_{-i}^i,~\mathrm{and}~\ell<L \\
w_p^j(0), & \mathrm{if}~j\neq i,~a_j\neq \hat{a}_j^i,~a_i=a_i^*,~a_k=\hat{a}_k^i,\forall k\neq i,j,~\ell<L \\
w_p^j(0), & \mathrm{if}~a_j\neq \hat{a}_j(i)~\mathrm{and}~\mathbf{a}_{-j}=\mathbf{\hat{a}}_{-j}(i),~\mathrm{and}~\ell=L \end{array}\right.. \nonumber
\end{eqnarray}
\end{itemize}

\subsection{Achieving Pure-action Payoff As an SPE Payoff}
We prove that for any pure-action payoff profile $\mathbf{v}=\mathbf{u}(\tilde{a}_0,\mathbf{\tilde{a}})$ that allows pure-action player-specific
punishments $\{\mathbf{v}^i=\mathbf{u}(\hat{a}_0(i),\mathbf{\hat{a}}(i))\}_{i=1}^N$, there exists $\underline{\delta}<1$ and a strategy profile
described in the above automaton, such that for all $\delta\in(\underline{\delta},1)$, the strategy profile is a subgame perfect equilibrium with
payoff $\mathbf{u}(\tilde{a}_0,\mathbf{\tilde{a}})$.

Since the payoff can be achieved by pure-action profile $(\tilde{a}_0,\mathbf{\tilde{a}})$, the equilibrium outcome path is repeating
$(\tilde{a}_0,\mathbf{\tilde{a}})$ in every period. Hence, the automaton is simplified to have $T=1$ and
$(\tilde{a}_0^0,\mathbf{\tilde{a}}^0)=(\tilde{a}_0,\mathbf{\tilde{a}})$.

Now we calculate the values of all the states in the automaton. For $w_e(0)$, we have
\begin{eqnarray}
\mathbf{V}(w_e(0))=(1-\delta)\cdot\mathbf{u}(\tilde{a}_0,\mathbf{\tilde{a}}) + \delta\cdot \mathbf{V}(w_e(0)) \Rightarrow
\mathbf{V}(w_e(0))=\mathbf{u}(\tilde{a}_0,\mathbf{\tilde{a}}).
\end{eqnarray}
For $w_p^i(L)$, we have
\begin{eqnarray}
\mathbf{V}(w_p^i(L))=(1-\delta)\cdot\mathbf{u}(\hat{a}_0(i),\mathbf{\hat{a}}(i)) + \delta\cdot \mathbf{V}(w_p^i(L)) \Rightarrow
\mathbf{V}(w_p^i(L))=\mathbf{u}(\hat{a}_0(i),\mathbf{\hat{a}}(i)).
\end{eqnarray}
For $w_p^i(\ell),~0\leq \ell<L-1$, we have
\begin{eqnarray}
\mathbf{V}(w_p^i(\ell)) &=& (1-\delta) \cdot \mathbf{u}(\hat{a}_0^i,a_i^*,\mathbf{\hat{a}}_{-i}^i)+\delta \cdot \mathbf{V}(w_p^i(\ell+1)) \\
&=& (1-\delta) \cdot \left[\mathbf{u}(\hat{a}_0^i,a_i^*,\mathbf{\hat{a}}_{-i}^i)\cdot\frac{1-\delta^{L-\ell-1}}{1-\delta}\right] + \delta^{L-\ell-1}
\cdot \mathbf{V}(w_p^i(L-1))
\end{eqnarray}
Since we can calculate $\mathbf{V}(w_p^i(L-1))$ as
\begin{eqnarray}
\mathbf{V}(w_p^i(L-1)) = (1-\delta)\cdot\mathbf{u}(\hat{a}_0^i,a_i^*,\mathbf{\hat{a}}_{-i}^i) + \delta\cdot\mathbf{V}(w_p^i(L)),
\end{eqnarray}
we have for $\ell=0,\ldots,L-1$,
\begin{eqnarray}
\mathbf{V}(w_p^i(\ell)) = (1-\delta^{L-\ell})\cdot\mathbf{u}(\hat{a}_0^i,a_i^*,\mathbf{\hat{a}}_{-i}^i) +
\delta^{L-\ell}\cdot\mathbf{u}(\hat{a}_0(i),\mathbf{\hat{a}}(i)).
\end{eqnarray}

With the values of the states, we can derive the conditions under which the strategy prescribed by this automaton is a subgame perfect equilibrium.
To this end, we need to check that for any state $w$ accessible from $w^0$, $f(w)$ is a Nash equilibrium of the normal-form game described by the
payoff function $g^w:\mathcal{A}\rightarrow\mathbb{R}^N$, where
\begin{eqnarray}
g^w(\mathbf{a})=(1-\delta)\cdot\mathbf{u}(a_0(w),\mathbf{a})+\delta\cdot\mathbf{V}(\lambda(w,a_0(w),\mathbf{a})),
\end{eqnarray}
where $a_0(w)$ is the intervention action prescribed in $f(w)$.

First, we check the incentive compatibility constraints for users to follow the action profile $\mathbf{\tilde{a}}$. This can be done by checking if
$\mathbf{\tilde{a}}$ is a Nash equilibrium of the normal-form game at the state $w=w_e(0)$. The normal-form game at state $w=w_e(0)$ has payoff
function
\begin{eqnarray}
g^w(\mathbf{a})=\left\{\begin{array}{ll} (1-\delta)\cdot\mathbf{u}(\tilde{a}_0,\mathbf{\tilde{a}})+\delta\cdot\mathbf{V}(w_e(0)) =
\mathbf{u}(\tilde{a}_0,\mathbf{\tilde{a}}), & \mathrm{if}~\mathbf{a} = \mathbf{\tilde{a}} \\
(1-\delta)\cdot\mathbf{u}(\tilde{a}_0,\mathbf{a})+\delta\cdot\mathbf{V}(w_p^i(0)), &
\mathrm{if}~a_i\neq\tilde{a}_i~\mathrm{and}~\mathbf{a}_{-i}=\mathbf{\tilde{a}}_{-i}
\end{array}\right..
\end{eqnarray}
The action profile $\mathbf{\tilde{a}}$ is a Nash equilibrium if and only if, for all $i\in\mathcal{N}$ and for all $a_i\in A_i$,
\begin{eqnarray}
u_i(\tilde{a}_0,\mathbf{\tilde{a}}) \geq (1-\delta)\cdot u_i(\tilde{a}_0,a_i,\mathbf{\tilde{a}_{-i}}) + \delta\cdot\left[(1-\delta^{L})\cdot
u_i(\hat{a}_0^i,a_i^*,\mathbf{\tilde{a}_{-i}}^i) + \delta^{L}\cdot u_i(\hat{a}_0(i),\mathbf{\hat{a}}(i))\right].
\end{eqnarray}
Define $M\triangleq\max_{i,a_0,\mathbf{a}} u_i(a_0,\mathbf{a})$. Then it suffices to have, for all $i\in\mathcal{N}$,
\begin{eqnarray}
v_i \geq (1-\delta)\cdot M + \delta\cdot\left[(1-\delta^{L})\cdot \underline{v}_i^w + \delta^{L}\cdot v_i^i\right].
\end{eqnarray}
After rearranging the terms, we have
\begin{eqnarray}\label{eqn:FolkTheorem_PlayerSpecific_delta_L_1}
\delta(1-\delta^L)\cdot(v_i-\underline{v}_i^w)+\delta^{L+1}\cdot(v_i-v_i^i) \geq (1-\delta)\cdot (M-v_i).
\end{eqnarray}
Note that for fixed $L$, the left hand side is strictly positive for any $\delta\in(0,1)$, and the right hand side goes to $0$ when
$\delta\rightarrow1$.

Second, we check the incentive compatibility constraints for staying in the punishment phase. For the states $w_p^i(0),\ldots,w_p^i(L-1)$, we check
if $(\hat{a}_0^i,a_i^*,\mathbf{\hat{a}}_{-i}^i$ is a Nash equilibrium of the corresponding normal-form games. The normal-form game at the state
$w=w_p^i(\ell)$ for $0\leq\ell\leq L-1$ has the payoff function
\begin{eqnarray}
g^w(\mathbf{a})=\left\{\begin{array}{ll}(1-\delta)\cdot\mathbf{u}(\hat{a}_0^i,\mathbf{a})+\delta\cdot\mathbf{V}(w_p^i(0)), & \mathrm{if}~a_i\neq
a_i^*,\mathbf{a}_{-i}=\mathbf{\hat{a}}_{-i}^i \\ (1-\delta)\cdot\mathbf{u}(\hat{a}_0^i,\mathbf{a})+\delta\cdot\mathbf{V}(w_p^j(0)), &
\mathrm{if}~j\neq i,a_j\neq \hat{a}_j^i,a_i=a_i^*,a_k=\hat{a}_k^i,\forall k\neq i,j \\ (1-\delta)\cdot\mathbf{u}(\hat{a}_0^i,\mathbf{a})+\delta\cdot\mathbf{V}(w_p^i(\ell+1)), & \mathrm{if}~\mathbf{a}=(a_i^*,\mathbf{\hat{a}}_{-i}^i) \\
\end{array}\right.. \nonumber
\end{eqnarray}
The action profile $(a_i^*,\mathbf{\hat{a}}_{-i}^i)$ is a Nash equilibrium if, for all $j\in\mathcal{N}$
\begin{eqnarray}
V_j(w_p^i(\ell)) \geq (1-\delta)\cdot M + \delta\cdot V_j(w_p^j(0)),
\end{eqnarray}
which is equivalent to
\begin{eqnarray}
&&(1-\delta^{L-\ell})\cdot u_j(\hat{a}_0^i,a_i^*,\mathbf{\hat{a}}_{-i}^i)+\delta^{L-\ell} u_j(\hat{a}_0(i),\mathbf{\hat{a}}(i)) \nonumber \\
&\geq& (1-\delta)\cdot M + \delta\cdot\left[(1-\delta^L)\cdot u_j(\hat{a}_0^j,a_j^*,\mathbf{\hat{a}}_{-j}^j)+\delta^L\cdot
u_j(\hat{a}_0(j),\mathbf{\hat{a}}(j))\right]. \nonumber
\end{eqnarray}
The above inequality can be further simplified as
\begin{eqnarray}
(1-\delta^{L-\ell})\cdot u_j(\hat{a}_0^i,a_i^*,\mathbf{\hat{a}}_{-i}^i)+\delta^{L-\ell}\cdot v_j^i \geq (1-\delta)\cdot M +
\delta\cdot\left[(1-\delta^L)\cdot \underline{v}_j^w+\delta^L\cdot v_j^j\right]. \nonumber
\end{eqnarray}
After rearranging the terms, we have
\begin{eqnarray}\label{eqn:FolkTheorem_PlayerSpecific_delta_L_2}
\delta^{L+1}\cdot(v_j^i-v_j^j) &\geq& (1-\delta)(M-u_j(\hat{a}_0^i,a_i^*,\mathbf{\hat{a}}_{-i}^i)) +
\delta(1-\delta^{L-\ell-1})(\underline{v}_j^w-u_j(\hat{a}_0^i,a_i^*,\mathbf{\hat{a}}_{-i}^i)) \nonumber \\
&+& \delta^{L-\ell}(1-\delta^{\ell+1})(\underline{v}_j^w-v_j^i),
\end{eqnarray}
which holds for a fixed $L$ when $\delta$ is large enough.

For the state $w_p^i(L)$, we check if $\mathbf{\hat{a}}(i)$ is a Nash equilibrium of the corresponding normal-form game. The normal-form game at the
state $w=w_p^i(\ell)$ for $0\leq\ell\leq L-1$ has the payoff function
\begin{eqnarray}
g^w(\mathbf{a})=\left\{\begin{array}{ll}(1-\delta)\cdot\mathbf{u}(\hat{a}_0(i),\mathbf{a})+\delta\cdot\mathbf{V}(w_p^j(0)), & \mathrm{if}~a_j\neq
\hat{a}_j(i),\mathbf{a}_{-j}=\mathbf{\hat{a}}_{-j}(i) \\ (1-\delta)\cdot\mathbf{u}(\hat{a}_0(i),\mathbf{a})+\delta\cdot\mathbf{V}(w_p^i(L)), & \mathrm{if}~\mathbf{a}=\mathbf{\hat{a}}(i) \\
\end{array}\right.. \nonumber
\end{eqnarray}
The action profile $\mathbf{\hat{a}}(i)$ is a Nash equilibrium if, for all $j\in\mathcal{N}$
\begin{eqnarray}
V_j(w_p^i(L)) \geq (1-\delta)\cdot M + \delta\cdot V_j(w_p^j(0)),
\end{eqnarray}
which is equivalent to
\begin{eqnarray}
v_j^i &\geq& (1-\delta)\cdot M + \delta\cdot\left[(1-\delta^L)\cdot u_j(\hat{a}_0^j,a_j^*,\mathbf{\hat{a}}_{-j}^j)+\delta^L\cdot
u_j(\hat{a}_0(j),\mathbf{\hat{a}}(j))\right] \nonumber \\
&=& (1-\delta)\cdot M + \delta\cdot\left[(1-\delta^L)\cdot \underline{v}_j^w+\delta^L\cdot v_j^j\right].
\end{eqnarray}
After rearranging the terms, we have
\begin{eqnarray}
\delta(1-\delta^L)\cdot(v_j-\underline{v}_j^w)+\delta^{L+1}\cdot(v_j^i-v_j^j) \geq (1-\delta)\cdot (M-v_j).
\end{eqnarray}
When $j\neq i$, we have
\begin{eqnarray}\label{eqn:FolkTheorem_PlayerSpecific_delta_L_3}
\delta(1-\delta^L)\cdot(v_j-\underline{v}_j^w)+\delta^{L+1}\cdot(v_j^i-v_j^j) \geq (1-\delta)\cdot (M-v_j),
\end{eqnarray}
which holds true for any fixed $L$ when $\delta\rightarrow1$ for the same reason as \eqref{eqn:FolkTheorem_PlayerSpecific_delta_L_2}. When $j=i$, we
have
\begin{eqnarray}\label{eqn:FolkTheorem_PlayerSpecific_delta_L_4}
(\delta+\cdots+\delta^L)\cdot(v_j-\underline{v}_j^w) \geq M-v_j.
\end{eqnarray}

Now we have four sets of constraints (\ref{eqn:FolkTheorem_PlayerSpecific_delta_L_1}), (\ref{eqn:FolkTheorem_PlayerSpecific_delta_L_2}),
(\ref{eqn:FolkTheorem_PlayerSpecific_delta_L_3}), and (\ref{eqn:FolkTheorem_PlayerSpecific_delta_L_4}) for the discount factor $\delta$ and the
length of punishment $L$. For any fixed $L$, \eqref{eqn:FolkTheorem_PlayerSpecific_delta_L_1} and \eqref{eqn:FolkTheorem_PlayerSpecific_delta_L_3}
hold true when $\delta\rightarrow1$, because the left hand sides of both inequalities are strictly positive and the right hand sides of both
inequalities go to $0$. For any fixed $L$, \eqref{eqn:FolkTheorem_PlayerSpecific_delta_L_2} also holds true when $\delta\rightarrow1$, because the
left hand side is strictly positive and the right hand side goes to $0$. For \eqref{eqn:FolkTheorem_PlayerSpecific_delta_L_4} to hold true, we choose
$L$ such that $L\cdot(v_j-\underline{v}_j^w) \geq M-v_j$ for all $i$. Then we can find a discount factor $\underline{\delta}^p<1$ such that all the
four sets of constraints are satisfied.

\subsection{Achieving Any Feasible and Individually Rational Payoff}
We prove that any feasible and individually rational payoff $\mathbf{v}\in\mathscr{V}_w^{\dag p}$ that allows player-specific punishments
$\{\mathbf{v}^i\}_{i=1}^N$ can be achieved as an SPE payoff. The difficulty is that starting from a certain period, the payoff on the equilibrium
path or the payoff in the player-specific punishments may be too low to prevent users from deviating at that period. This difficulty is resolved by
\cite[Lemma 3.7.2]{MailathSamuelson}, which states that for any $\varepsilon>0$, there exists $\underline{\delta}^\prime<1$ (resp.
$\underline{\delta}^{\prime i}<1$) such that for any payoff $\mathbf{v}\in\mathscr{V}_w^{\dag p}$ (resp. $\mathbf{v}^i$) and any discount factor
$\delta\in(\underline{\delta}^\prime,1)$ (resp. $\delta\in(\underline{\delta}^{\prime i},1)$), there exists a sequence of pure action profiles that
gives discounted average payoff $\mathbf{v}$ (resp. $\mathbf{v}^i$) and that has continuation payoffs within $\varepsilon$ of $\mathbf{v}$ (resp.
$\mathbf{v}^i$) at any period $t$. In other words, the payoff starting from any period is approximately the same, as if generated by a pure-action
profile. Combining with the results in the previous subsection, there exists
$\underline{\delta}=\max\{\underline{\delta}^p,\underline{\delta}^\prime,\max_{i\in\mathcal{N}} \underline{\delta}^{\prime i}\}$, such that any
discount factor $\delta\in(\underline{\delta},1)$ can sustain $\mathbf{v}\in\mathscr{V}_w^{\dag p}$ as an SPE payoff.

\section{Proof of Theorem~1}\label{appendix:MinimumDiscountFactor_EquilibriumPayoffSet}
We state the outline of the proof first. The proof heavily replies on the concept of self-generating sets. Simply put, a self-generating set,
associated with a discount factor, is a set in which every payoff is an SPE payoff under the associated discount factor.\footnote{We refer
interesting readers to \cite[Section~2.5.1]{MailathSamuelson} for the definition of self-generating sets.} Any self-generating set has a minimum
discount factor associated with it; any discount factor that is larger than the minimum one can be associated with that self-generating set. The idea
of the proof is to find the ``optimal'' self-generating set, the one with the smallest associated minimum discount factor, among all the
self-generating sets that include the target payoff. In order to find such a self-generating set, we first derive the minimum discount factor
associated with a self-generating set. Then we minimize the minimum discount factor associated with a self-generating set, over all the
self-generating sets that include the target payoff. Note that for the sake of analytical tractability, we confine our search to a special class of
self-generating sets, which is the reason why we obtain the upper bound of the minimum discount factor to support a target payoff, instead of the
minimum discount factor itself.

\subsection{Minimum Discount Factor to Support Self-generating Sets}
First, we calculate the minimum discount factor associated with a self-generating set, which can be very difficult for an arbitrary self-generating
set. To obtain analytical results, we consider the self-generating sets of the following form:
\begin{eqnarray}
\mathscr{Y}_{\mu} = \left\{\mathbf{v}\in\mathscr{V}_w^{\dag p}:\sum_{i=1}^N \frac{v_i}{\bar{v}_i}=1,~v_i\geq\mu_i,~\forall
i\in\mathcal{N}\right\}\bigcup\left\{\mathbf{\underline{v}}^w\right\},
\end{eqnarray}
where $\mathbf{\underline{v}}^w$ is the minmax payoff profile with the $i$th element being $\underline{v}_i^w$. For any target payoff on the Pareto
boundary of the feasible set of the protocol design problem, we can find a self-generating set $\mathscr{Y}_{\mu}$ defined as above to include it.

According to \cite[Definition~2.5.3]{MailathSamuelson}, a set $\mathscr{Y}_{\mu}$ is self-generating if every payoff in $\mathscr{Y}_{\mu}$ is
pure-action decomposable on $\mathscr{Y}_{\mu}$. A payoff $\mathbf{v}$ is pure-action decomposable on $\mathscr{Y}_{\mu}$, if there exists a pure
action profile $(a_0^*,\mathbf{a}^*)$ and a specification of continuation promises $\gamma: \mathcal{A}\rightarrow\mathscr{Y}_{\mu}$, such that for
all $i\in\mathcal{N}$ and for all $a_i\in A_i$,
\begin{eqnarray}
v_i=(1-\delta)u_i(a_0^*,\mathbf{a}^*) + \delta\gamma_i(\mathbf{a}^*) \geq (1-\delta)u_i(a_0^*,a_i,\mathbf{a}_{-i}^*) +
\delta\gamma_i(a_i,\mathbf{a}_{-i}^*).
\end{eqnarray}

By Assumption~1, the mutual minmax payoff profile $\mathbf{\underline{v}}^w$ is a Nash equilibrium payoff profile. Hence, regardless of the discount
factor, $\mathbf{\underline{v}}^w$ can be pure-action decomposed by the mutual minmax profile $(\hat{a}_0,\mathbf{\hat{a}})$ and the continuation
promise $\gamma(\mathbf{a})=\mathbf{\underline{v}}^w$ for all $\mathbf{a}\in\mathcal{A}$.

For any payoff profile $\mathbf{v}\in\mathscr{Y}_{\mu}\setminus\mathbf{\underline{v}}^w$, we first prove that $\mathbf{v}$ must be decomposed by a
pure action in $\{(\underline{a}_0,\mathbf{\check{a}}^1),\ldots,(\underline{a}_0,\mathbf{\check{a}}^N)\}$, and then find out the conditions on the
discount factor under which $\mathbf{v}$ is pure-action decomposable on $\mathscr{Y}_{\mu}$.
\begin{lemma}
Any payoff profile $\mathbf{v}\in\mathscr{Y}_{\mu}$ other than $\mathbf{\underline{v}}^w$ must be decomposed by a pure action in
$\{(\underline{a}_0,\mathbf{\check{a}}^1),\ldots,(\underline{a}_0,\mathbf{\check{a}}^N)\}$.
\end{lemma}
\begin{IEEEproof}
From $\mathbf{v}=(1-\delta)\mathbf{u}(a_0^*,\mathbf{a}^*) + \delta\gamma(\mathbf{a}^*)$, we know that $\mathbf{v}$ is the convex combination of
$\mathbf{u}(a_0^*,\mathbf{a}^*)$ and $\gamma(\mathbf{a}^*)$. Since $\mathbf{v}$ and $\gamma(\mathbf{a}^*)$ are both in $\mathscr{Y}_{\mu}$, we have
\begin{eqnarray}
\sum_{i=1}^N \frac{v_i}{\bar{v}_i} = \sum_{i=1}^N \frac{\gamma(\mathbf{a}^*)}{\bar{v}_i} = 1.
\end{eqnarray}
Hence, we have $\sum_{i=1}^N (u_i(a_0^*,\mathbf{a}^*)/\bar{v}_i) = 1$. The pure actions that satisfy this condition must come from the set
$\{(\underline{a}_0,\mathbf{\check{a}}^1),\ldots,(\underline{a}_0,\mathbf{\check{a}}^N)\}$.
\end{IEEEproof}

Suppose $\mathbf{v}$ is decomposed by $(\underline{a}_0,\mathbf{\check{a}}^i)$. Let $y_{ij}$ be the maximum payoff that user $j$ can get by deviating
from $(\underline{a}_0,\mathbf{\check{a}}^i)$, namely
\begin{eqnarray}
y_{ij}=\max_{a_j} u_j(\underline{a}_0, a_j, \mathbf{\check{a}}_{-j}^i).
\end{eqnarray}
Then the constraint on pure-action decomposability can be written as
\begin{eqnarray}
v_j = (1-\delta) u_j(\underline{a}_0,\mathbf{\check{a}}^i) + \delta \gamma_j(\mathbf{\check{a}}^i) \geq (1-\delta)
u_j(\underline{a}_0,a_j,\mathbf{\check{a}}_{-j}^i) + \delta \gamma_j(a_j,\mathbf{\check{a}}_{-j}^i),~\forall j\in\mathcal{N}.
\end{eqnarray}
As long as we set $\gamma_i(\mathbf{a})=\gamma_i(\mathbf{\check{a}}^i)$ for all $\mathbf{a}\in\mathcal{A}$, user $i$ will have no incentive to
deviate from $(\underline{a}_0,\mathbf{\check{a}}^i)$. Therefore, we only need to consider $j\neq i$. To get the minimum discount factor, we let
$\gamma_j(\mathbf{a})=\underline{v}_j^w$ for all $\mathbf{a}\neq\mathbf{\check{a}}^i$. Then the above constraint simplifies to
\begin{eqnarray}
v_j = (1-\delta)\cdot0 + \delta \gamma_j(\mathbf{\check{a}}^i) \geq (1-\delta) y_{ij} + \delta \underline{v}_j^w,~\forall j\neq i,
\end{eqnarray}
which is equivalent to
\begin{eqnarray}\label{eqn:DiscountFactor_IncentiveConstraint}
\delta \geq \frac{y_{ij}-v_j}{y_{ij}-\underline{v}_j^w},~\forall j\neq i.
\end{eqnarray}

Besides the above incentive constraint, the discount factor should ensure that there exists $\gamma(\mathbf{\check{a}}^i)\in\mathscr{Y}_{\mu}$ such
that $\mathbf{v}=(1-\delta) \mathbf{u}(\underline{a}_0,\mathbf{\check{a}}^i) + \delta \gamma(\mathbf{\check{a}}^i)$. In other words, we need
\begin{eqnarray}
\mu_i\leq\gamma_i(\mathbf{\check{a}}^i))=\bar{v}_i-\frac{\bar{v}_i-v_i}{\delta}\leq\bar{v}_i~\Leftrightarrow~\frac{\bar{v}_i-v_i}{\bar{v}_i-\mu_i}\leq\delta<1,
\end{eqnarray}
and
\begin{eqnarray}
\mu_j\leq\gamma_j(\mathbf{\check{a}}^i))=\frac{v_j}{\delta}\leq\bar{v}_j~\Leftrightarrow~\frac{v_j}{\bar{v}_j}\leq\delta\leq\frac{v_i}{\mu_i},~\forall
j\neq i.
\end{eqnarray}
Because $\frac{\bar{v}_i-v_i}{\bar{v}_i-\mu_i}\geq\frac{\bar{v}_i-v_i}{\bar{v}_i}=\sum_{j\neq i} \frac{v_j}{\bar{v}_j}\geq \frac{v_j}{\bar{v}_j}$ for
any $j\neq i$, the above two constraints simplify to
\begin{eqnarray}\label{eqn:DiscountFactor_ExistenceContinuationPayoff}
\delta\geq\frac{\bar{v}_i-v_i}{\bar{v}_i-\mu_i}.
\end{eqnarray}
Combining (\ref{eqn:DiscountFactor_IncentiveConstraint}) and (\ref{eqn:DiscountFactor_ExistenceContinuationPayoff}), the minimum discount factor
under which $\mathbf{v}$ is pure-action decomposable by $(\underline{a}_0,\mathbf{\check{a}^i})$ on $\mathscr{Y}_{\mu}$ is
\begin{eqnarray}
\max\left\{\frac{\bar{v}_i-v_i}{\bar{v}_i-\mu_i},~\max_{j\neq i} \frac{y_{ij}-v_j}{y_{ij}-\underline{v}_j^w}\right\}.
\end{eqnarray}

For a given payoff $\mathbf{v}$, we should choose an action profile from
$\{(\underline{a}_0,\mathbf{\check{a}}^1),\ldots,(\underline{a}_0,\mathbf{\check{a}}^N)\}$, so that the discount factor is minimized. Specifically,
the minimum discount factor under which $\mathbf{v}$ is decomposed on $\mathscr{Y}_{\mu}$ is
\begin{eqnarray}
\min_{i\in\mathcal{N}}\max\left\{\frac{\bar{v}_i-v_i}{\bar{v}_i-\mu_i},~\max_{j\neq i} \frac{y_{ij}-v_j}{y_{ij}-\underline{v}_j^w}\right\}.
\end{eqnarray}
As a result, the minimum discount factor under which $\mathscr{Y}_{\mu}$ is self-generating is
\begin{eqnarray}\label{eqn:MinimumDiscountFactor_SelfGenerating_Formulation}
\delta_{\mu}=\max_{\mathbf{v}\in\mathscr{Y}_{\mu}}\min_{i\in\mathcal{N}}\max\left\{\frac{\bar{v}_i-v_i}{\bar{v}_i-\mu_i},~\max_{j\neq i}
\frac{y_{ij}-v_j}{y_{ij}-\underline{v}_j^w}\right\},
\end{eqnarray}
which can be obtained analytically according to the following lemma.

\begin{lemma}\label{lemma:MinimumDiscountFactor_SelfGenerating}
The minimum discount factor under which $\mathscr{Y}_{\mu}$ is self-generating, defined in
(\ref{eqn:MinimumDiscountFactor_SelfGenerating_Formulation}), is
\begin{eqnarray}\label{eqn:MinimumDiscountFactor_SelfGenerating}
\delta_{\mu}=\max\left\{\frac{N-1}{N-\sum_{i=1}^N \mu_i/\bar{v}_i},~\max_{i\in\mathcal{N}}\max_{j\neq i}
\frac{y_{ij}-\mu_j}{y_{ij}-\underline{v}_j^w}\right\}.
\end{eqnarray}
\end{lemma}
\begin{IEEEproof}
To simplify the notation, we define a matrix $\mathbf{X}(\mathbf{v})\in\mathbb{R}^{N\times N}$, whose diagonal elements are defined as
\begin{eqnarray}
x_{ii}(\mathbf{v}) = \frac{\bar{v}_i-v_i}{\bar{v}_i-\mu_i},~\forall i\in\mathcal{N},
\end{eqnarray}
and whose off-diagonal elements are defined as
\begin{eqnarray}
x_{ij}(\mathbf{v}) = \frac{y_{ij}-v_j}{y_{ij}-\underline{v}_j^w},~\forall i\in\mathcal{N}, j\neq i.
\end{eqnarray}
Note that one important property of $\mathbf{X}(\mathbf{v})$ is
\begin{eqnarray}\label{eqn:PropertyX}
x_{jj}(\mathbf{v})=\frac{\bar{v}_j-v_j}{\bar{v}_j-\mu_j}>\frac{y_{ij}-v_j}{y_{ij}-\mu_j}\geq\frac{y_{ij}-v_j}{y_{ij}-\underline{v}_j^w}=x_{ij}(\mathbf{v}),~\forall
i\neq j~\mathrm{and}~\forall \mathbf{v}.
\end{eqnarray}

The optimization problem in (\ref{eqn:MinimumDiscountFactor_SelfGenerating_Formulation}) is equivalent to
\begin{eqnarray}\label{eqn:MinimumDiscountFactor_SelfGenerating_Formulation_X}
\delta_{\mu} = \max_{\mathbf{v}\in\mathscr{Y}_{\mu}} \min_{i\in\mathcal{N}} \max_{j\in\mathcal{N}} x_{ij}(\mathbf{v}).
\end{eqnarray}
The optimal value $\delta_{\mu}$ must be strictly smaller than $1$. This is because $\delta_{\mu}=1$ only if there exists a $\mathbf{v}$ such that
$\max_{j\in\mathcal{N}} x_{ij}(\mathbf{v})=1$ for all $i\in\mathcal{N}$, which is possible only if $v_i=\mu_i$ for all $i\in\mathcal{N}$. But such a
$\mathbf{v}$ is not in $\mathscr{Y}_{\mu}$, because $\sum_{i=1}^N (\mu_i/\bar{v}_i)<1$.

Without compromising optimality, we consider the solution $\mathbf{v}^*$ to the optimization problem
\eqref{eqn:MinimumDiscountFactor_SelfGenerating_Formulation_X} as one of the following two types:
\begin{itemize}
\item Type-1 solutions: at the optimal solution $\mathbf{v}^*$, $x_{ii}(\mathbf{v}^*)=\delta_{\mu}$ for all $i\in\mathcal{N}$.
\item Type-2 solutions: at the optimal solution $\mathbf{v}^*$, there exists $i$ and $j~(i\neq j)$ such that $\delta_{\mu}=x_{ij}(\mathbf{v}^*)$, but there exists no $i$ such that
$\delta_{\mu}=x_{ii}(\mathbf{v}^*)$.
\end{itemize}

The reason why we only need to consider the above two types is as follows. Assume that at the optimal solution $\mathbf{v}^*$, there exists
$i^*\in\mathcal{N}$, such that $\delta_{\mu}=x_{i^* i^*}(\mathbf{v}^*)$. Under this assumption, we claim that unless $x_{i
i}(\mathbf{v}^*)=\delta_{\mu}$ for all $i\in\mathcal{N}$, we can always find another solution $\mathbf{v}^\prime$ such that $\min_{i\in\mathcal{N}}
\max_{j\in\mathcal{N}} x_{ij}(\mathbf{v}^\prime)\geq\min_{i\in\mathcal{N}} \max_{j\in\mathcal{N}} x_{ij}(\mathbf{v}^*)$ and $x_{i^*
i^*}(\mathbf{v}^\prime)>\delta_{\mu}$. As a result, we only need to consider the solution $\mathbf{v}^*$ with $x_{ii}(\mathbf{v}^*)=\delta_{\mu}$ for
all $i\in\mathcal{N}$ or $\mathbf{v}^*$ with $x_{ii}(\mathbf{v}^*)\neq\delta_{\mu}$ for any $i\in\mathcal{N}$.

Define the set $\mathcal{I}=\{i\in\mathcal{N}:x_{ii}(\mathbf{v}^*)=\delta_{\mu}\}$. In the following, we prove our claim that if $\mathcal{I}$ is
nonempty and is a strict subset of $\mathcal{N}$ ($\mathcal{N}\setminus\mathcal{I}$ is nonempty), we can always find another solution
$\mathbf{v}^\prime$ such that $\min_{i\in\mathcal{N}} \max_{j\in\mathcal{N}} x_{ij}(\mathbf{v}^\prime)\geq\min_{i\in\mathcal{N}}
\max_{j\in\mathcal{N}} x_{ij}(\mathbf{v}^*)$ and $x_{ii}(\mathbf{v}^\prime)>\delta_{\mu}$ for $i\in\mathcal{I}$.
\begin{itemize}
\item Suppose that there exists $i^\prime\in\mathcal{N}$, such that $x_{i^\prime i^\prime}(\mathbf{v}^*)<\delta_{\mu}$. Then define a new payoff profile
$\mathbf{v}^\prime\in\mathscr{Y}_{\mu}$, with $v_i^\prime=v_i^*-\varepsilon\cdot\bar{v}_i$ for all $i\in\mathcal{I}$ \footnote{We can always find a
small enough $\varepsilon$ such that $v_i^\prime>\mu_i$, because $v_i^*>\mu_i$, which results from the fact that $x_{i^*
i^*}(\mathbf{v}^*)=\delta_{\mu}<1$.} and $v_{i^\prime}^\prime=v_{i^\prime}^*+|\mathcal{I}|\cdot\varepsilon\cdot\bar{v}_{i^\prime}$, where
$\varepsilon>0$ is small enough such that $v_i^\prime\in(\mu_i,\bar{v}_i)$ for all $i\in\mathcal{N}$. Since $x_{j i^\prime}(\mathbf{v}^*)<x_{i^\prime
i^\prime}(\mathbf{v}^*)<\delta_{\mu}$ for all $j\neq i^\prime$ and $\max_{j\in\mathcal{N}} x_{ij}(\mathbf{v}^*)\geq \delta_{\mu}$ for all
$i\in\mathcal{N}$, we know that $x_{i i^\prime}(\mathbf{v}^*)<\max_{j} x_{ij}(\mathbf{v}^*)$ for all $i\in\mathcal{N}$. Hence, increasing
$v_{i^\prime}^*$ to $v_{i^\prime}^\prime$ does not lower the maximum over each row of the matrix $\mathbf{X}$, namely $\max_{j\in\mathcal{N}}
x_{ij}(\mathbf{v}^\prime)\geq\max_{j\in\mathcal{N}} x_{ij}(\mathbf{v}^*)$ for all $i\in\mathcal{N}$. By construction, we have
$x_{ii}(\mathbf{v}^\prime)>\delta_{\mu}$ for $i\in\mathcal{I}$.
\item Suppose that $x_{ii}(\mathbf{v}^*)>\delta_{\mu}$ for all $i\in\mathcal{N}\setminus\mathcal{I}$. Then pick an arbitrary
$i^\prime\in\mathcal{N}\setminus\mathcal{I}$, and define a new payoff profile $\mathbf{v}^\prime\in\mathscr{Y}_{\mu}$, with
$v_i^\prime=v_i^*-\varepsilon\cdot\bar{v}_i$ for all $i\in\mathcal{I}$ and
$v_{i^\prime}^\prime=v_{i^\prime}^*+|\mathcal{I}|\cdot\varepsilon\cdot\bar{v}_{i^\prime}$, where $\varepsilon>0$ is small enough such that
$v_i^\prime\in(\mu_i,\bar{v}_i)$ for all $i\in\mathcal{N}$, and $\max_{i\in\mathcal{I}}
x_{ii}(\mathbf{v}^\prime)<\min_{i\in\mathcal{N}\setminus\mathcal{I}} x_{ii}(\mathbf{v}^\prime)$. In this way, we still have
$x_{ii}(\mathbf{v}^\prime)=\max_{j} x_{ij}(\mathbf{v}^\prime)$ for $i\in\mathcal{I}$, and have $x_{ii}(\mathbf{v}^\prime)=\min_{k}\max_{j}
x_{kj}(\mathbf{v}^\prime)$ for $i\in\mathcal{I}$. However, $x_{ii}(\mathbf{v}^\prime)>x_{ii}(\mathbf{v}^*)=\delta_{\mu}$ for $i\in\mathcal{I}$, which
contradicts the assumption that $\mathbf{v}^*$ is the optimal solution.
\end{itemize}

Now we have proved that if $\mathcal{I}$ is nonempty, we only need to consider type-1 solutions. Otherwise (when $\mathcal{I}$ is empty), we consider
type-2 solutions. In the following, we solve the optimization problem (\ref{eqn:MinimumDiscountFactor_SelfGenerating_Formulation_X}) by considering
the solutions of the above two types.

For a type-1 solution $\mathbf{v}^1$, since $\mathcal{I}=\mathcal{N}$, we have
\begin{eqnarray}
\frac{\bar{v}_i-v_i^1}{\bar{v}_i-\mu_i} = c,~\forall i\in\mathcal{N}.
\end{eqnarray}
Using $\sum_{i=1}^N (v_i^1/\bar{v}_i) = 1$, we can solve $c$ as
\begin{eqnarray}\label{eqn:OptimalValue_Type1Solution}
c = \frac{N-1}{N-\sum_{i=1}^N (\mu_i/\bar{v}_i)}.
\end{eqnarray}

For a type-2 solution $\mathbf{v}^2$, the optimal value is upper bounded by
\begin{eqnarray}
\max_{i\in\mathcal{N}} \max_{j\neq i} \frac{y_{ij}-\mu_j}{y_{ij}-\underline{v}_j^w} \triangleq
\frac{y_{i^*j^*}-\mu_{j^*}}{y_{i^*j^*}-\underline{v}_{j^*}^w}.
\end{eqnarray}
One $\mathbf{v}'$ that can possibly achieve the upper bound is
\begin{eqnarray}
v_i'=\mu_i,~\forall i\neq i^*~\mathrm{and}~v_{i^*}'=\bar{v}_{i^*}(1-\sum_{j\neq i^*} \mu_j/\bar{v}_j).
\end{eqnarray}
Then $\max_{j} x_{ij}(\mathbf{v}') = 1$ for all $i\neq i^*$. In other words, $\min_{i} \max_{j} x_{ij}(\mathbf{v}') = \max_{j}
x_{i^*j}(\mathbf{v}')$. Since $x_{i^*j^*}(\mathbf{v}')=\frac{y_{i^*j^*}-\mu_{j^*}}{y_{i^*j^*}-\underline{v}_{j^*}^w}\geq x_{i^*j}(\mathbf{v}')$ for
all $j\neq i^*$, $x_{i^*j^*}(\mathbf{v}')=\max_{j} x_{i^*j}(\mathbf{v}')$ if and only if
\begin{eqnarray}
x_{i^*j^*}(\mathbf{v}')\geq x_{i^*i^*}(\mathbf{v}') = \frac{\bar{v}_{i^*}-v_{i^*}'}{\bar{v}_{i^*}-\mu_{i^*}}.
\end{eqnarray}
Since $v_{i^*}'\geq v_i$ for any $\mathbf{v}\in\mathscr{Y}_{\mu}$, we have $x_{i^*i^*}(\mathbf{v}')\leq c$, where $c$ is the optimal value of type-1
solutions in (\ref{eqn:OptimalValue_Type1Solution}). As a result, if $x_{i^*j^*}(\mathbf{v}')< x_{i^*i^*}(\mathbf{v}')$, the optimal value of type-2
solutions must be smaller than that of type-1 solutions. Then there is no need to study type-2 solutions when $x_{i^*j^*}(\mathbf{v}')<
x_{i^*i^*}(\mathbf{v}')$. When $x_{i^*j^*}(\mathbf{v}')\geq x_{i^*i^*}(\mathbf{v}')$, the optimal value achieved by type-2 solutions is
\begin{eqnarray}
x_{i^*j^*}(\mathbf{v}') = \max_{i\in\mathcal{N}} \max_{j\neq i} \frac{y_{ij}-\mu_j}{y_{ij}-\underline{v}_j^w}.
\end{eqnarray}

Finally, the optimal value of (\ref{eqn:MinimumDiscountFactor_SelfGenerating_Formulation}) is the maximum of the optimal values of type-1 and type-2
solutions, which is expressed as in (\ref{eqn:MinimumDiscountFactor_SelfGenerating}).
\end{IEEEproof}

\subsection{Minimum Discount Factor to Support The Target Payoff}
Now we can calculate the minimum discount factor to support the target payoff $\mathbf{v}^\star$. As we have discussed at the beginning of the proof,
this is equivalent to find the minimum discount to support a self-generating set that includes the target payoff. Since we restrict our search in a
particular class of self-generating sets $\mathscr{Y}_{\mu}$, the discount factor obtained in this way is an upper bound of the minimum discount
factor. This upper bound can be written explicitly as
\begin{eqnarray}\label{eqn:MinimumDiscountFactor_TargetPayoffSet_Formulation}
\bar{\delta}(\mathbf{v}^\star) = \displaystyle\min_{\mu}
\delta_{\mu},~\mathrm{subject~to}~\mathbf{v}^\star\in\mathscr{Y}_{\mu}\setminus\{\mathbf{\underline{v}}^w\}.
\end{eqnarray}
According to Lemma~\ref{lemma:MinimumDiscountFactor_SelfGenerating}, $\delta_{\mu}$ can be calculated using
\eqref{eqn:MinimumDiscountFactor_SelfGenerating} as
\begin{eqnarray}
\delta_{\mu}=\max\left\{\frac{N-1}{N-\sum_{i=1}^N \mu_i/\bar{v}_i},~\max_{i\in\mathcal{N}}\max_{j\neq i}
\frac{y_{ij}-\mu_j}{y_{ij}-\underline{v}_j^w}\right\}. \nonumber
\end{eqnarray}
In the rest of this proof, we solve the above optimization problem to obtain $\bar{\delta}(\mathbf{v}^\star)$.

First, observe that
\begin{eqnarray}
\max_{i\in\mathcal{N}}\max_{j\neq i} \frac{y_{ij}-\mu_j}{y_{ij}-\underline{v}_j^w}=\max_{j\in\mathcal{N}}\max_{i\neq j}
\frac{y_{ij}-\mu_j}{y_{ij}-\underline{v}_j^w}\leq1.
\end{eqnarray}
Defining $w_j = \max_{i\neq j} y_{ij}$, we can rewrite the optimization problem \eqref{eqn:MinimumDiscountFactor_TargetPayoffSet_Formulation} as
\begin{eqnarray}\label{eqn:MinimumDiscountFactor_TargetPayoffSet_Reformulation}
\bar{\delta}(\mathbf{v}^\star) = &\displaystyle\min_{\mu}& \max\left\{\frac{N-1}{N-\sum_{i=1}^N \mu_i/\bar{v}_i},~\max_{j\in\mathcal{N}}
\frac{w_j-\mu_j}{w_j-\underline{v}_j^w}\right\} \\
& s.t. & \underline{v}_i^w\leq\mu_i\leq v_i^\star,~\forall i\in\mathcal{N}. \nonumber
\end{eqnarray}

Then we prove that one of the (possibly many) optimal solutions to (\ref{eqn:MinimumDiscountFactor_TargetPayoffSet_Reformulation}) should satisfy
\begin{eqnarray}
\frac{w_j-\mu_j^*}{w_j-\underline{v}_j^w} = C < 1,~\forall j\in\mathcal{N}.
\end{eqnarray}
First, as long as $\mu$ is such that $\underline{v}_i^w<\mu_i<v_i^\star$ for all $i\in\mathcal{N}$, we have
\begin{eqnarray}
\frac{N-1}{N-\sum_{i=1}^N \mu_i/\bar{v}_i}<\frac{N-1}{N-\sum_{i=1}^N v_i^\star/\bar{v}_i}=1,~\mathrm{and}~
\frac{w_j-\mu_j}{w_j-\underline{v}_j^w}<\frac{w_j-\underline{v}_j^w}{w_j-\underline{v}_j^w}=1~\forall j\in\mathcal{N}.
\end{eqnarray}
Hence, any optimal solution $\mu^*$ should satisfy
\begin{eqnarray}
\frac{w_j-\mu_j^*}{w_j-\underline{v}_j^w} < 1,~\forall j\in\mathcal{N}.
\end{eqnarray}
Suppose that at the optimal solution $\mathbf{\mu}^*$, there is a nonempty set $\mathcal{J}=\{j\in\mathcal{N}:
\frac{w_j-\mu_j^*}{w_j-\underline{v}_j^w}<\max_{i\in\mathcal{N}} \frac{w_i-\mu_i^*}{w_i-\underline{v}_i}\}$. Then we define a new self-generating set
$\mathscr{Y}_{\mu^\prime}$, with $\mu_j^\prime$ satisfying $\frac{w_j-\mu_j'}{w_j-\underline{v}_j^w}=\max_{i\in\mathcal{N}}
\frac{w_i-\mu_i^*}{w_i-\underline{v}_i}$ for all $j\in\mathcal{J}$, and with $\mu_j'=\mu_j^*$ for all $j\notin\mathcal{J}$. Note that $\mu_j'$
satisfies $\underline{v}_j^w<\mu_j'<\mu_j^*$ for $j\in\mathcal{J}$, which implies $\frac{N-1}{N-\sum_{i=1}^N
\mu_i'/\bar{v}_i}<\frac{N-1}{N-\sum_{i=1}^N \mu_i^*/\bar{v}_i}$. Consequently, we have
\begin{eqnarray}
\max\left\{\frac{N-1}{N-\sum_{i=1}^N \mu_i'/\bar{v}_i},~\max_{j\in\mathcal{N}} \frac{w_{j}-\mu_j'}{w_{j}-\underline{v}_j^w}\right\}\leq
\max\left\{\frac{N-1}{N-\sum_{i=1}^N \mu_i^*/\bar{v}_i},~\max_{j\in\mathcal{N}} \frac{w_{j}-\mu_j^*}{w_{j}-\underline{v}_j^w}\right\}, \nonumber
\end{eqnarray}
which gives us an optimal solution $\mu^\prime$ whose corresponding $\mathcal{J}$ is empty. In other words, there exists an optimal solution $\mu^*$
such that
\begin{eqnarray}\label{eqn:ConditionOptimalSelfGeneratingSet}
\frac{w_j-\mu_j^*}{w_j-\underline{v}_j^w}=\max_{i\in\mathcal{N}} \frac{w_i-\mu_i^*}{w_i-\underline{v}_i^w} \triangleq C,~\forall j\in\mathcal{N}.
\end{eqnarray}
Note that $\max_{j\in\mathcal{N}} \frac{w_j-v_j^\star}{w_j-\underline{v}_j^w}\leq C\leq 1$ due to the constraints on $\mu_j$.

Since we can write $\mu_j^*$ as $\mu_j^*=w_j-(w_j-\underline{v}_j^w)\cdot C$, the optimization problem
(\ref{eqn:MinimumDiscountFactor_TargetPayoffSet_Reformulation}) can be further simplified into an optimization problem with one decision variable
$C$:
\begin{eqnarray}\label{eqn:MinimumDiscountFactor_TargetPayoffSet_Reformulation_SingleVariable}
\bar{\delta}(\mathbf{v}^\star) = &\displaystyle\min_{C}& \max\left\{\frac{N-1}{N-\sum_{i=1}^N ((w_i-(w_i-\underline{v}_i^w)\cdot C)/\bar{v}_i)},~C\right\} \\
& s.t. & \max_{j\in\mathcal{N}} \frac{w_j-v_j^\star}{w_j-\underline{v}_j^w}\leq C\leq 1. \nonumber
\end{eqnarray}

Since $\frac{N-1}{N-\sum_{i=1}^N ((w_i-(w_i-\underline{v}_i^w)\cdot C)/\bar{v}_i)}$ is decreasing in $C$, the optimal $C^*$ should satisfy
\begin{eqnarray}\label{eqn:OptimalC}
\frac{N-1}{N-\sum_{i=1}^N ((w_i-(w_i-\underline{v}_i)\cdot C^*)/\bar{v}_i)}=C^*,
\end{eqnarray}
if $\frac{N-1}{N-\sum_{i=1}^N ((w_i-(w_i-\underline{v}_i)\cdot C^*)/\bar{v}_i)}\geq\max_{j\in\mathcal{N}}
\frac{w_j-v_j^\star}{w_j-\underline{v}_j^w}$. The only solution to (\ref{eqn:OptimalC}) that is smaller than $1$ can be calculated as
$\frac{2(N-1)}{(N-T)+\sqrt{(N-T)^2+4(N-1)(T-S)}}$, where $T=\sum_{i=1}^N (w_i/\bar{v}_i)$ and $S=\sum_{i=1}^N (\underline{v}_i^w/\bar{v}_i)$.

Finally, we obtain an upper bound of the minimum discount factor to sustain $\mathbf{v}^\star$ from solving
(\ref{eqn:MinimumDiscountFactor_TargetPayoffSet_Reformulation_SingleVariable}):
\begin{eqnarray}
\bar{\delta}(\mathbf{v}^\star) = \max\left\{\max_{i\in\mathcal{N}}
\frac{w_i-v_i^\star}{w_i-\underline{v}_i^w},~\frac{2(N-1)}{(N-T)+\sqrt{(N-T)^2+4(N-1)(T-S)}}\right\}
\end{eqnarray}

\section{Proof of Theorem~2}\label{appendix:EquilibriumStrategy}
In the algorithm in Table~I, we first determine the self-generating set $\mathscr{Y}_{\nu}$ under the discount factor
$\bar{\delta}(\mathbf{v}^\star)$ that includes the target payoff $\mathbf{v}^\star$. From the previous appendix, we know that the self-generating set
$\mathscr{Y}_{\nu}$ can be of the following form
\begin{eqnarray}
\mathscr{Y}_{\nu} = \left\{\mathbf{v}\in\mathscr{V}_w^{\dag p}: \sum_{i=1}^N \frac{v_i}{\bar{v}_i} = 1,~v_i\geq \nu_i,~\forall
i\in\mathcal{N}\right\}\bigcup\left\{\mathbf{\underline{v}}\right\}.
\end{eqnarray}
We also know that such a self-generating set $\mathscr{Y}_{\nu}$ exists, and satisfies \eqref{eqn:ConditionOptimalSelfGeneratingSet}, namely
\begin{eqnarray}
\frac{w_j-\nu_j}{w_j-\underline{v}_j^w}=\max_{i\in\mathcal{N}} \frac{w_i-\nu_i}{w_i-\underline{v}_i^w} \triangleq C,~\forall j\in\mathcal{N}.
\end{eqnarray}
As long as we can find $C$, we can determine $\mathscr{Y}_{\nu}$ by $\nu_i=w_i-(w_i-\underline{v}_i)\cdot C$ for all $i\in\mathcal{N}$. According to
the second part of the previous appendix, we know that $C=\bar{\delta}(\mathbf{v}^\star)$. Hence, we set
$\nu_i=w_i-(w_i-\underline{v}_i)\cdot\bar{\delta}(\mathbf{v}^\star)$ in the initialization.

Since $\mathscr{Y}_{\nu}$ is self-generating under the discount factor $\bar{\delta}(\mathbf{v}^\star)$, it is self-generating under the given
discount factor $\delta\geq\bar{\delta}(\mathbf{v}^\star)$ according to \cite[Proposition~7.3.4]{MailathSamuelson}. Since the target payoff
$\mathbf{v}^\star$ is in the self-generating set $\mathscr{Y}_{\nu}$, it can be decomposed by a pure action $\mathbf{\tilde{a}}^0$ and a continuation
payoff $\mathbf{v}(1)\in\mathscr{Y}_{\nu}$. According to the definition of self-generation, the continuation payoff $\mathbf{v}(1)$ can also be
decomposed by a pure action $\mathbf{\tilde{a}}^1$ and a continuation payoff $\mathbf{v}(2)$. Repeating this procedure, we can obtain the desired
sequence of pure-action profiles $\{\mathbf{\tilde{a}}^\tau\}_{\tau=0}^{T-1}$.

\newpage

%\begin{figure}
%\centering
%\includegraphics[width =3.5in]{AdHoc_TwoUser_color}
%\caption{An example power control game with two users. The distance from user 1's transmitter to its receiver is normalized to $1$, and the distance
%from user 2's transmitter to its receiver is $0.5$. The vertical distance between the two users' transmitters and that between the two users'
%receivers are both $0.5$.} \label{fig:AdHoc_TwoUser_color}
%\end{figure}
%
%\begin{figure}
%\centering
%\includegraphics[width =4.5in]{UtilityRegion_PowerControl_2user}
%\caption{Payoffs in a two-user power control game with different values of the maximum intervention power. The network topology is shown in
%Fig.~\ref{fig:AdHoc_TwoUser_color}. We assume that the channel gain $h$ is reciprocal to the distance $d$ with the path loss exponent $3$, that is,
%$h=d^{-3}$. The noise powers at the receivers of both users are $0.2$ watts. The power budgets of both users are $1$ watts. The dash curve is the
%Pareto boundary of the set of pure-action payoffs. The solid line is the Pareto boundary of the set of feasible payoffs. The circle is the mutual
%minmax payoff in each case. The gray area is the set of feasible and individually rational payoffs. The asterisk is the payoff at Nash equilibrium of
%the stage game without intervention.} \label{fig:UtilityRegion_PowerControl_2user}
%\end{figure}
%
%\begin{figure}
%\centering
%\includegraphics[width =3.5in]{FlowControl_Diagram}
%\caption{System diagram for flow control.} \label{fig:FlowControl_Diagram}
%\end{figure}

\begin{figure}
\centering
\includegraphics[width =6.0in]{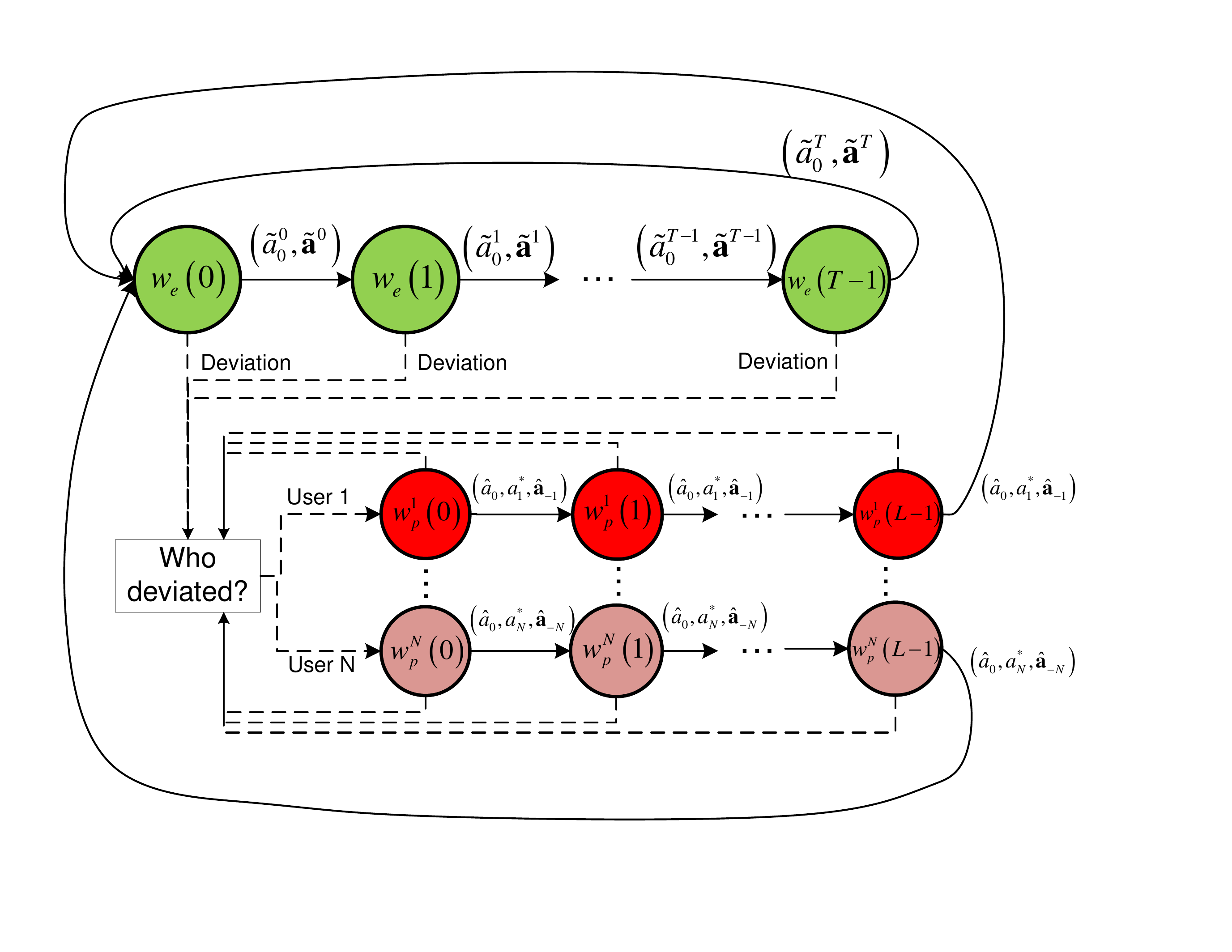}
\caption{The automaton of the equilibrium strategy of the game with a mutual minmax profile $(\hat{a}_0,\mathbf{\hat{a}})$. Circles are states, where
$\left\{w_e(\tau)\right\}_{\tau=0}^{T-1}$ is the set of states in the equilibrium outcome path, and $\left\{w_p^i(\ell)\right\}_{\ell=0}^{L-1}$ is
the set of states in the punishment phase for user $i$. The initial state is $w_e(0)$. Solid arrows are the prescribed state transitions labeled by
the action profiles leading to the transitions. Dashed arrows are the state transitions when deviation happens. $a_i^*$ is user $i$'s best response
to $\hat{a}_0$ and $\mathbf{\hat{a}_{-i}}$.} \label{fig:Automata_RepeatedGames_NotNE}
\end{figure}

\begin{table}
\renewcommand{\arraystretch}{1.1}
\caption{Algorithm to Generate $\{\mathbf{\tilde{a}}^\tau\}_{\tau=0}^{T-1}$, Users' Action Profiles In The Equilibrium Outcome Path.}
\label{table:EquilibriumStrategy} \centering
% Some packages, such as MDW tools, offer better commands for making tables
% than the plain LaTeX2e tabular which is used here.
% \begin{tabularx}{0.5\textwidth}{|c|X|X|X|X|}
%\begin{tabularx}{\textwidth}{l}
\begin{tabular}{l}
\hline
\textbf{Require:} The target payoff $\mathbf{v}^\star \in \mathscr{P}_{\gamma}$ and the discount factor $\delta \geq \bar{\delta}(\mathbf{v}^\star)$ \\
\hline
\textbf{Initialization:} Set $\nu_i=w_i-(w_i-\underline{v}_i)\cdot\delta$ for all $i\in\mathcal{N}$, $\tau=0$, $\mathbf{v}(0)=\mathbf{v}^\star$. \\
\textbf{Repeat} \\
~~~~find $i^*$ such that $\frac{1}{\delta}v_j(\tau)-\frac{1-\delta}{\delta}\cdot u_j(\underline{a}_0,\mathbf{\check{a}}^{i^*}) \geq \nu_j$ for all
$j\in\mathcal{N}$ \\
~~~~$\mathbf{v}(\tau+1)=\frac{1}{\delta}\mathbf{v}(\tau)-\frac{1-\delta}{\delta}\cdot\mathbf{u}(\underline{a}_0,\mathbf{\check{a}}^{i^*})$ \\
~~~~$\mathbf{\tilde{a}}^\tau = \mathbf{\check{a}}^{i^*}$ \\
~~~~$\tau\leftarrow \tau+1$ \\
\textbf{Until} $\mathbf{v}(\tau)=\mathbf{v}^{\star}$ \\
\hline
%\end{tabularx}
\end{tabular}
\end{table}

\begin{table*}
\renewcommand{\arraystretch}{1.1}
\caption{Performance Comparison Among Different Schemes Under Different Minimum Payoffs Guarantees \newline(discount factors for repeated games shown
in the parenthesis)} \label{table:PerformanceComparison} \centering
% Some packages, such as MDW tools, offer better commands for making tables
% than the plain LaTeX2e tabular which is used here.
% \begin{tabularx}{0.5\textwidth}{|c|X|X|X|X|}
\begin{tabularx}{\textwidth}{|c|c|c|c|c|c|}
\hline
Min. payoff & Metrics & NE \cite{BharathKumarJaffe}--\cite{ZhangDouligeris_TCOM} & One-shot \cite{ParkMihaela_JSAC}\cite{GaiKrishnamachari_Infocom}\cite{YiMihaela_TCOM} & Repeated w/o intervention & Repeated with intervention \\
\hline
\multirow{2}{*}{$\gamma_i=1$} & Sum Payoff & 39.3 & 110.4 & 110.2 (1.000) & 114.2 (0.987) \\ & Absolute Fairness & 4.0 & 9.6 & 16.7 (0.861) & 16.7 (0.840) \\
\hline
\multirow{2}{*}{$\gamma_i=3$} & Sum Payoff & 39.3 & 85.8 & 108.2 (1.000) & 108.2 (0.962) \\ & Absolute Fairness & 4.0 & 10.6 & 16.7 (0.861) & 16.7 (0.840) \\
\hline
\multirow{2}{*}{$\gamma_i=7$} & Sum Payoff & N/A & 64.4 & 96.2 (0.960) & 96.2 (0.910) \\ & Absolute Fairness & N/A & 10.3 & 16.7 (0.861) & 16.7 (0.840) \\
\hline
\multirow{2}{*}{$\gamma_i=14$} & Sum Payoff & N/A & N/A & 75.2 (0.861) & 75.2 (0.840) \\ & Absolute Fairness & N/A & N/A & 16.7 (0.861) & 16.7 (0.840) \\
\hline
\end{tabularx}
\end{table*}

\end{document}